\documentclass{emulateapj}
\usepackage{txfonts}
\usepackage{graphicx}
\usepackage{subfigure}
\usepackage{multirow}
\usepackage{verbatim}
\usepackage{url}

\def\lsim{\mathrel{\rlap{\lower4pt\hbox{\hskip1pt$\sim$}}
    \raise1pt\hbox{$<$}}}                
\def\gsim{\mathrel{\rlap{\lower4pt\hbox{\hskip1pt$\sim$}}
    \raise1pt\hbox{$>$}}}                

\shorttitle{Three New Eclipsing White-dwarf - M-dwarf Binaries}
\shortauthors{N.M. Law et al.}

\begin{document}

\title{Three New Eclipsing White-dwarf - M-dwarf Binaries Discovered in a Search for Transiting Planets Around M-dwarfs}

\author{Nicholas M. Law\altaffilmark{1}, Adam L. Kraus\altaffilmark{2}, Rachel Street\altaffilmark{3}, Benjamin J. Fulton\altaffilmark{3},  Lynne A. Hillenbrand\altaffilmark{4}, Avi Shporer\altaffilmark{3,5}, Tim Lister\altaffilmark{3}, Christoph Baranec\altaffilmark{4}, Joshua S. Bloom\altaffilmark{6}, Khanh Bui\altaffilmark{4}, Mahesh P. Burse\altaffilmark{7}, S. Bradley Cenko\altaffilmark{6}, H.K. Das\altaffilmark{7}, Jack. T.C. Davis\altaffilmark{4}, Richard G. Dekany\altaffilmark{4}, Alexei V. Filippenko\altaffilmark{6}, Mansi M. Kasliwal\altaffilmark{4}, S. R. Kulkarni\altaffilmark{4}, Peter Nugent\altaffilmark{8}, Eran O. Ofek\altaffilmark{4}, Dovi Poznanski\altaffilmark{9}, Robert M. Quimby\altaffilmark{4}, A. N. Ramaprakash\altaffilmark{7}, Reed Riddle\altaffilmark{4}, Jeffrey M. Silverman\altaffilmark{6}, Suresh Sivanandam\altaffilmark{1}, Shriharsh Tendulkar\altaffilmark{4}}

\altaffiltext{1}{Dunlap Fellow, Dunlap Institute for Astronomy and Astrophysics, University of Toronto, 50 St. George Street, Toronto M5S 3H4, Ontario, Canada}
\altaffiltext{2}{Hubble Fellow, Institute for Astronomy, University of Hawaii, 2680 Woodlawn Drive, Honolulu, HI, 96822, USA}
\altaffiltext{3}{Las Cumbres Observatory Global Telescope Network, Inc., 6740 Cortona Dr. Suite 102, Santa Barbara, CA 93117 USA}
\altaffiltext{4}{Cahill Center for Astrophysics, California Institute of Technology, Pasadena, CA, 91125, USA}
\altaffiltext{5}{Department of Physics, Broida Hall, University of California, Santa Barbara, CA 93106, USA}
\altaffiltext{6}{Department of Astronomy, University of California, Berkeley, CA 94720-3411, USA}
\altaffiltext{7}{Inter-University Centre for Astronomy \& Astrophysics, Ganeshkhind, Pune, 411007, India}
\altaffiltext{8}{Computational Cosmology Center, Lawrence Berkeley National Laboratory, 1 Cyclotron Road, Berkeley, CA 94720, USA}
\altaffiltext{9}{School of Physics and Astronomy, Tel-Aviv University, Tel Aviv 69978, Israel}

\begin{abstract}
We present three new eclipsing white-dwarf / M-dwarf binary systems discovered during a search for transiting planets around M-dwarfs. Unlike most known eclipsing systems of this type, the optical and infrared emission is dominated by the M-dwarf components, and the systems have optical colors and discovery light curves consistent with being Jupiter-radius transiting planets around early M-dwarfs. We detail the PTF/M-dwarf transiting planet survey, part of the Palomar Transient Factory (PTF). We present a Graphics Processing Unit (GPU)-based box-least-squares search for transits that runs approximately 8$\times$ faster than similar algorithms implemented on general purpose systems. For the discovered systems, we decompose low-resolution spectra of the systems into white-dwarf and M-dwarf components, and use radial velocity measurements and cooling models to estimate masses and radii for the white dwarfs. The systems are compact, with periods between 0.35 and 0.45 days and semimajor axes of approximately $\rm2 R_{\odot}$ (0.01 AU). The M-dwarfs have masses of approximately 0.35$\rm{M_{\odot}}$, and the white dwarfs are all hydrogen-atmosphere with temperatures of around 8000K, and have masses of approximately 0.5$\rm{M_{\odot}}$. We use the Robo-AO laser guide star adaptive optics system to tentatively identify one of the objects as a triple system. We also use high-cadence photometry to put an upper limit on the white dwarf radius of 0.025$\rm{R_{\odot}}$ (95\% confidence) in one of the systems. Accounting for our detection efficiency and geometric factors, we estimate that $\rm 0.08\%^{+0.10\%}_{-0.05\%}$ (90\% confidence) of M-dwarfs are in these short-period, post-common-envelope white-dwarf / M-dwarf binaries where the optical light is dominated by the M-dwarf. Similar eclipsing binary systems can have arbitrarily small eclipse depths in red bands and generate plausible small-planet-transit light curves. As such, these systems are a source of false positives for M-dwarf transiting planet searches. We present several ways to rapidly distinguish these binaries from transiting planet systems.
\end{abstract}

\keywords{}

\maketitle

\section{Introduction}
Large numbers of non-eclipsing white-dwarf / main-sequence binaries have been discovered in the Sloan Digital Sky Survey and other surveys (e.g. \citealt{Rebassa2011}; \citealt{Bianchi2007} and references therein). For low-mass stars in particular there is a bridge in color between white dwarfs and M-dwarfs. The bridge is interpreted as being due to rare white-dwarf / M-dwarf binaries, at a ratio with respect to single stars of $\sim$1:2300 \citep{Smolcic2004}. 

White-dwarf / M-dwarf \textit{eclipsing} systems are much rarer, and almost all have been discovered by searching for white dwarfs displaying very deep eclipses of up to several magnitudes (e.g. \citealt{Drake2010}). These searches find systems containing relatively hot ($>12000$K) white dwarfs and mid-to-late M-dwarfs. The discovery rate of these systems (e.g. \citealt{Drake2010, Parsons2011, Parsons2011b}) is increasing with the advent of large sky surveys. These binaries survived the common-envelope phase of their evolution and many will become cataclysmic variables (e.g. \citealt{Nebot2011}), and so the properties and number statistics of these systems can provide windows into two important areas of stellar evolution. Precision measurements of the systems allow the determination of the masses and radii of two types of stars for which there are relatively few measurements \citep{Nebot2009, Kraus2011, Pyrzas2011, Pyrzas2009}. 

In this paper we present three eclipsing white-dwarf/M-dwarf systems discovered during the PTF/M-dwarfs search for transiting planets around M-dwarfs. In contrast to most known eclipsing systems of this type, the systems detected in this survey have optical and infrared emission dominated by the M-dwarf component and contain relatively low-temperature (8000K) white dwarfs and relatively early M-dwarfs. The shape of the light curves of the detected systems is similar to that expected for transiting giant planets around M-dwarfs, in particular in having a flat-bottomed eclipse with a depth of 1-20\% in red optical bands.

The PTF/M-dwarfs survey \citep{Law2011} is a search for transiting planets around 100,000 M-dwarfs. The survey is performed with the Palomar Transient Factory (PTF) camera \citep{Rahmer2008, Law2009, Law2010} on the 48-inch Samuel Oschin telescope at Palomar Observatory, and is a Key Project of the Palomar Transient Factory \citep{Law2009, Rau2009}. The PTF/M-dwarfs survey is designed to complement other M-dwarf transiting planet surveys such as MEarth (e.g. \citealt{Charbonneau2009, Irwin2010}), the WFCam transit survey \citep{Sipocz2011} and the \mbox{M-dwarfs} in the Kepler mission target list \citep{Borucki2011}, by covering a much larger number of M-dwarfs at somewhat lower sensitivity. The survey achieves photometric precisions of a few percent for $\approx$100,000 targets, and few-millimag precision around a subset of $\approx$10,000 M-dwarfs. These systems offer much larger transit depths compared to solar-type stars, while their very red colors compared to most other stars in the field greatly reduce the probability of a blended eclipse producing a difficult-to-detect transit false positive.

\begin{figure*}
  \centering
  \resizebox{1.0\textwidth}{!}
   {
	\includegraphics{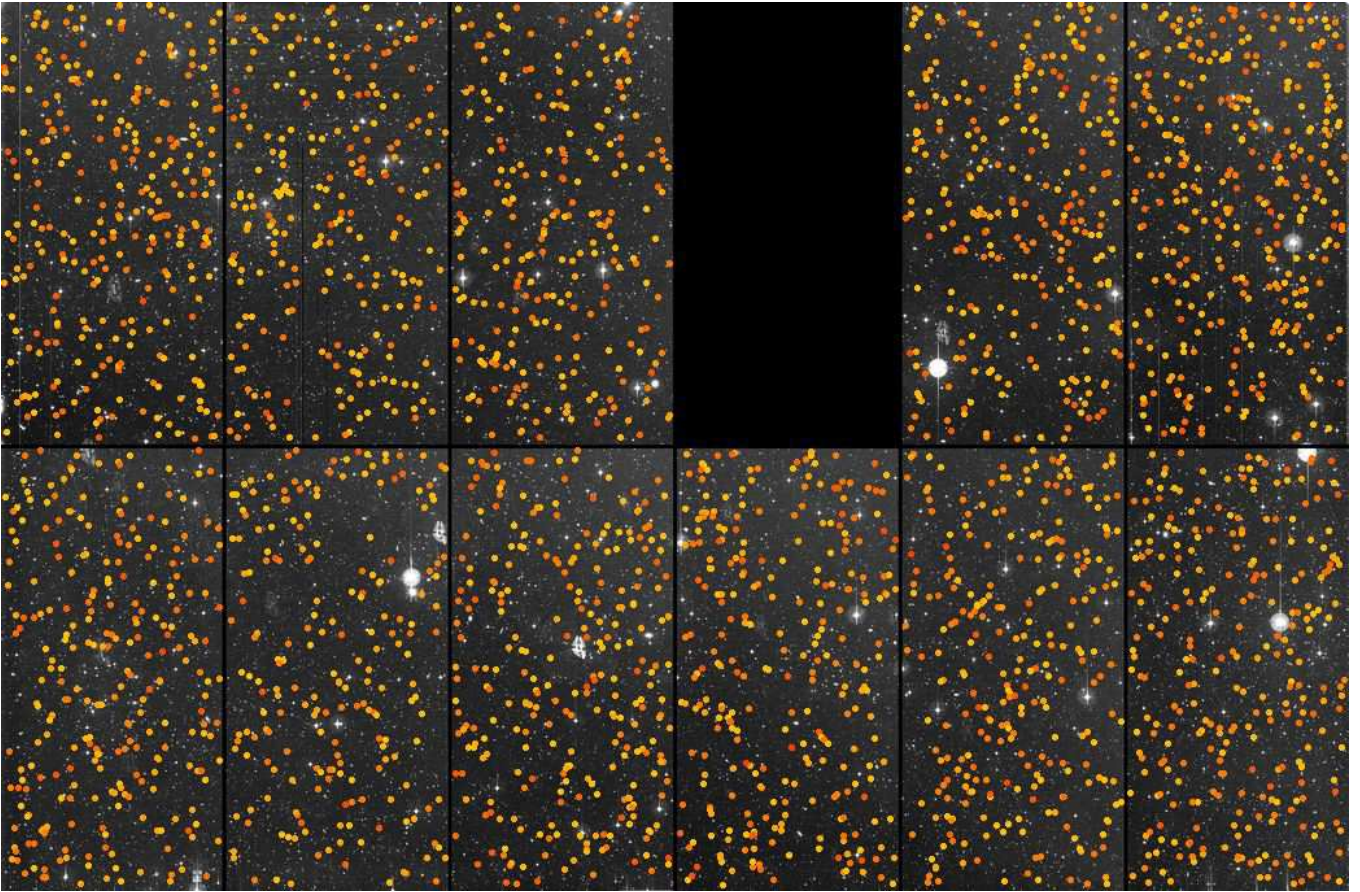}
   }
   \caption{A PTF Camera image of a survey field centered at 17:28, +57:22 and covering $3.50^{\circ}\times2.31^{\circ}$. The highlighted points show the 2,851 stars with photometrically-estimated spectral types later than K4 and photometric stability better than 5\%. The colors of the points correspond to the stellar temperatures, with K4 as yellow and late-M-dwarfs as red. The pipeline has removed stars with possible photometric precision problems such as proximity to a bright star or a bright ghost image. North is up and East is to the right.}

   \label{fig:field}
\end{figure*}

The three eclipsing systems presented here were originally detected as Jupiter-sized planet candidates during the first year of operations of the survey. Follow up of the candidates showed large color changes during eclipse and very large radial velocity signals, suggesting a hidden hot companion. In this paper we detail the properties of these eclipsing M-dwarf - white-dwarf systems and explore ways to distinguish them from true planetary transits.

The paper is organized as follows: in section \ref{sec:survey} we describe the PTF/M-dwarfs survey, its precision photometry methods, and its target detection strategies, including a new method of performing a box-least-squares transit search in parallel on GPU hardware. In section \ref{sec:disc_and_followup} we detail the three new eclipsing white-dwarf - M-dwarf systems and describe follow up photometric, low-resolution spectroscopic, and radial velocity observations, which are used to determine masses and radii for the system components in section \ref{sec:models}. Section \ref{sec:discussion} determines the frequency of eclipsing binaries such as these and discusses ways to distinguish them from transiting planets.
 
\section{The PTF/M-dwarfs Survey}
\label{sec:survey}

The PTF/M-dwarfs project consists of a transiting planet survey on the robotic 48-inch Samuel Oschin telescope (hereafter P48), photometric follow-up using the Palomar 60-inch telescope, the Byrne Observatory at Sedgwick Reserve (hereafter BOS), the LCOGT Faulkes-North and Faulkes-South telescopes (FTN \& FTS), and radial velocity follow-up with the HIRES instrument on Keck I. 

The 8-square-degree camera on the P48 telescope allows the survey to cover $\approx$3,000 M-dwarfs in each pointing (figure \ref{fig:field}), for a total of around 100,000 targets per year at galactic latitudes of $20-35^{\circ}$. The PTF/M-dwarfs survey typically observes several fields with an approximately 20-minute cadence for 4-5 hours per night. Individual fields are typically observed for several months, and observations are performed throughout the year. All PTF/M-dwarfs data is acquired with a 60-second exposure time in the PTF camera's Mould-R filter (similar to the Sloan Digital Sky Survey r filter). Data-taking is interleaved with the Palomar Transient Factory supernova survey which generally operates on 1-3 days cadences, and so scheduling constraints lead to a variety of final cadences for PTF/M-dwarfs fields. Table \ref{tab:survey_specs} summarizes the specifications of the PTF camera and the PTF/M-dwarfs survey.

The PTF/M-dwarfs survey is sensitive to equal-mass binaries around all M-dwarfs in the survey fields brighter than $\rm{m_R}$$\approx$20. In practise, the mass and radius determination and follow-up of such faint systems is extremely challenging, so we impose a discovery magnitude limit of $\rm{m_R}$$\approx$18. At that magnitude, the typical P48 data photometric precision is a few percent per datapoint, allowing immediate high-precision constraints on the system properties in discovery data. Given the saturation and faint cutoff limits of the survey, its effective distance ranges are \mbox{200-1300 pc} for M0 dwarfs, \mbox{50-290 pc} for M5 dwarfs, and \mbox{10-70 pc} for M9 dwarfs.

\begin{table}
\caption{\label{tab:survey_specs}The specifications of the PTF Camera and the PTF/M-dwarfs survey.}

\begin{tabular}{ll}
\multicolumn{2}{l}{\bf P48 PTF camera specifications }\\
\hline
Telescope     & Palomar 48-inch (1.2m) Samuel Oschin \\
Camera field dimensions & 3.50 $\times$ 2.31 degrees\\
Camera field of view        & 8.07 square degrees \\
Light sensitive area        & 7.26 square degrees \\
Image quality        & 2.0 arcsec FWHM in median seeing \\
Filters              & $\rm{g^\prime}$ \& Mould-R; other bands available\\
CCD specs            & 2K$\times$4K MIT/LL 3-edge butted CCDs\\
Plate scale             &  1.01 arcsec / pixel\\
Readout noise        & $<$ 12 $\rm{e^-}$\\
Readout speed        & 35 seconds, entire 100 MPix array\\
\vspace{0.25cm}\\

\multicolumn{2}{l}{\bf PTF/M-dwarfs survey characteristics}\\
\hline

Targets & Late-K, M and L dwarfs with $\rm{m_R} < 18$\\
Survey sky area & 29 square degrees every 2 months\\
Target locations & 20$^{\circ}$ $<$ galactic latitude $<$ $35^{\circ}$ \\
Targets covered & $\sim$12,000 every 2 months\\
Observations per night & 5 hours\\
Exposure time   & 60 seconds \\
Cadence & 15-25 minutes\\
Observation length & 1--3 months \\
Efficiency           & 66\% open-shutter (slew during readout)\\
Saturation Level     & $\rm{m_R} \sim 14$ (seeing dependant, in 60s)\\
Sensitivity (median)          & $\rm{m_R}$$\approx$21 in 60 s, 5$\sigma$\\
                               & $\rm{m_{g^\prime}}$$\approx$21.3 in 60 s, 5$\sigma$ \\
Photometric stability & 3 millimag (brightest targets) \\
                      & 10\% (faintest targets)\\
Followup & Photometric: Palomar 60-inch, FTN/S, BOS\\
         & Low-res spectroscopic: Lick Shane-3m\\
         & Radial velocity: Keck I / HIRES \\
\hline \vspace{0.25cm} \\
\end{tabular}
\end{table}

\subsection{P48 Data Reduction}
\label{sec:p48_reduction}
Reduction of the PTF/M-dwarfs data is performed in two steps: the standard PTF
realtime data reduction software first calibrates the images, and then a custom pipeline performs source-extraction, association and precision photometry. 

Immediately after observations, PTF data is transferred to Lawrence Berkeley National Laboratory where crosstalk corrections are applied to each chip, standard bias/overscan subtraction is performed, and a superflat based on recently acquired data is applied.

After the calibrated data is transferred to the Dunlap Institute for Astronomy and Astrophysics at University of Toronto, the PTF/M-dwarfs pipeline extracts sources from the calibrated images, produces an optimal photometric solution, associates sources in images taken at different times, and applies a range of eclipse-detection algorithms to the resulting light curves. We summarize the system
here; the pipeline is described in more detail in \citet{Law2011}.

Initial source extraction is performed on individual CCD chip images by SExtractor \citep{Bertin1996} using radius-optimized aperture photometry with a locally-optimized background. The extracted sources are filtered to remove those close to bad pixels, diffraction spikes from bright stars, and those that may be affected by nearby sources. Heliocentric Julian dates (HJD) are used for time measurements throughout the pipeline.

 The photometric zeropoints for each epoch are initially estimated
 based on either SDSS or USNO-B1 photometry for bright stars in the
 field.  The pipeline then optimizes the \mbox{zeropoint} of each epoch to
 minimize the median photometric variability of all the remaining
 sources. This first optimization typically improves the long-term
 photometric stability to below the percent level. The pipeline then
 filters the generated light curves, searching for epochs that
 produce anomalous photometry for a large fraction of the sources;
 those epochs are usually those affected by clouds, moonlight or some
 other effect that varies across the images. Typically 0-2\% of
 epochs are flagged by this process, and are removed from further
 consideration. A second iteration of variable-source removal and
 zeropoint optimization is then performed. The final zeropoints are
 applied to each light curve, along with flags for poor conditions,
 nearby sources that could cause confusion, bad pixels, and other problems that could affect the photometry. Running on a 2.5 GHz quad-core desktop computer the pipeline processes a 300-epoch set of 11 chips (54 GB of image data) in less than 24 hours.

\begin{figure}
  \centering
  \resizebox{1.0\columnwidth}{!}
   {
	\includegraphics{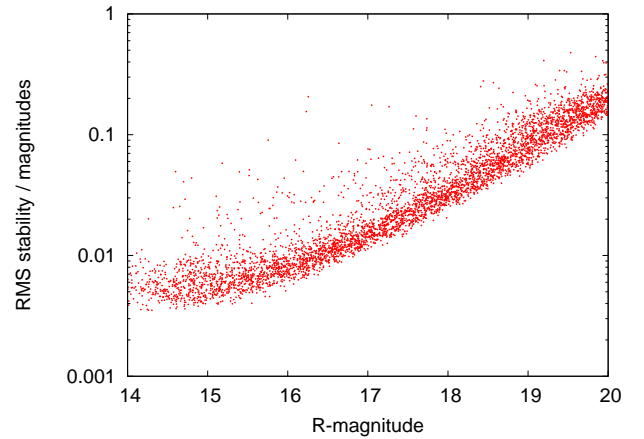}
   }
   \caption{The photometric stability achieved in a typical PTF/M-dwarfs field in 140 epochs spanning 92 days of observations. Each point corresponds to the RMS variability of one of the 23,713 stars in this field with magnitude $14<\rm{m_R}<20$. For clarity, 50\% of the fainter stars have been removed from this plot. The best stability achieved in this field is 4-millimags, with some CCD chips and sky regions having somewhat lower stability at 5-7 millimags. The few stars with very high variability compared to others at similar magnitudes are astrophysically varying sources such as eclipsing binaries, RR Lyrae stars and other variables. 
}

   \label{fig:phot_perf}
\end{figure}

The pipeline typically achieves a photometric stability of 3-5 millimags over periods of months (figure \ref{fig:phot_perf}). The photometric precision is photon-limited for all sources fainter than mR $\approx$16, except in regions of crowding or nebulosity. The pipeline has been used for several PTF projects such as the open cluster rotation project \citep{Agueros2011}.

\subsection{Eclipsing Binary Detection}
\label{eclipse_detection}
The pipeline produces light curves for 25,000-100,000 stars per field.  The large number of light curves makes the processing time involved in searching for eclipses an important consideration. An initial cut is made on the basis of the estimated spectral type of the source, based on photometry from the USNO-B1, 2MASS and (where available) Sloan Digital Sky Survey data (\citealt{Monet2003, Skrutskie2006, York2000}). The photometric data is fit to updated versions of the spectral energy distributions given in \citet{Kraus2007a}, giving an accuracy of approximately 1 subclass. Sources with late-K, M or L estimated spectral types are passed to our eclipse search algorithms.

\subsubsection{High-variability source searches}

Many eclipsing binaries have eclipse signals in the
tens-of-percent range as well as large eclipse duty cycles. These systems have significantly increased
photometric variability compared to nearby stars of similar
magnitude, and so a simple variability search can rapidly find
them (figure \ref{fig:phot_perf}). We estimate the value
and scatter of the locally expected photometric stability as a function of stellar magnitude in 0.1-mag bins, using a sigma-clipped average of the stars detected on each chip. Objects that show a more than $2\sigma$ increased
variability compared to the ensemble expectation are flagged for
further review. Of the systems presented here only PTFEB11.441 was detected in this manner; the photometric variations from the smaller eclipse depths of the other systems required more computationally-intensive algorithms.

\begin{figure*}[]
  \centering
	\subfigure{\resizebox{0.31\textwidth}{!}{{\includegraphics{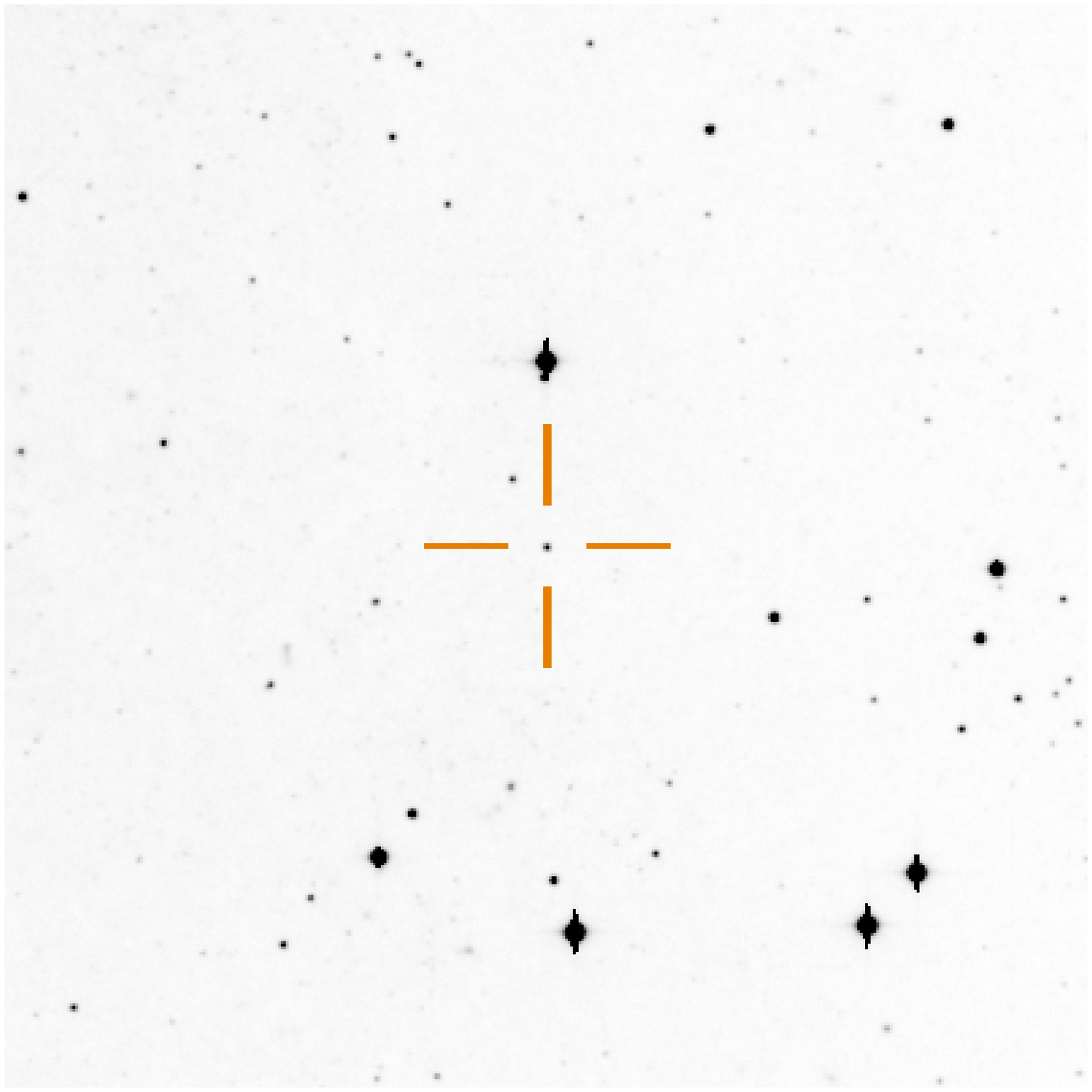}}}}\hspace{0.15in}
	\subfigure{\resizebox{0.31\textwidth}{!}{{\includegraphics{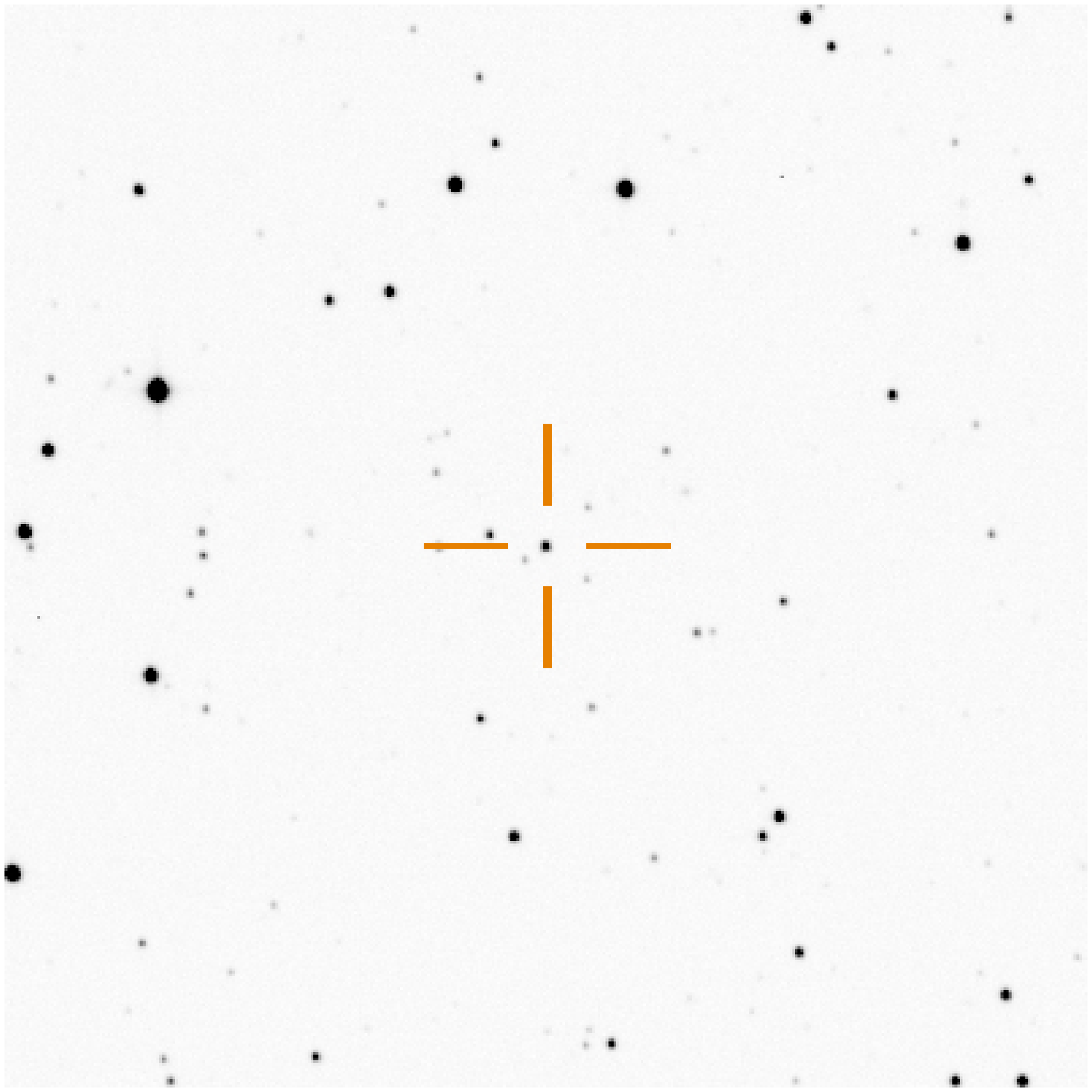}}}}\hspace{0.15in}
	\subfigure{\resizebox{0.31\textwidth}{!}{{\includegraphics{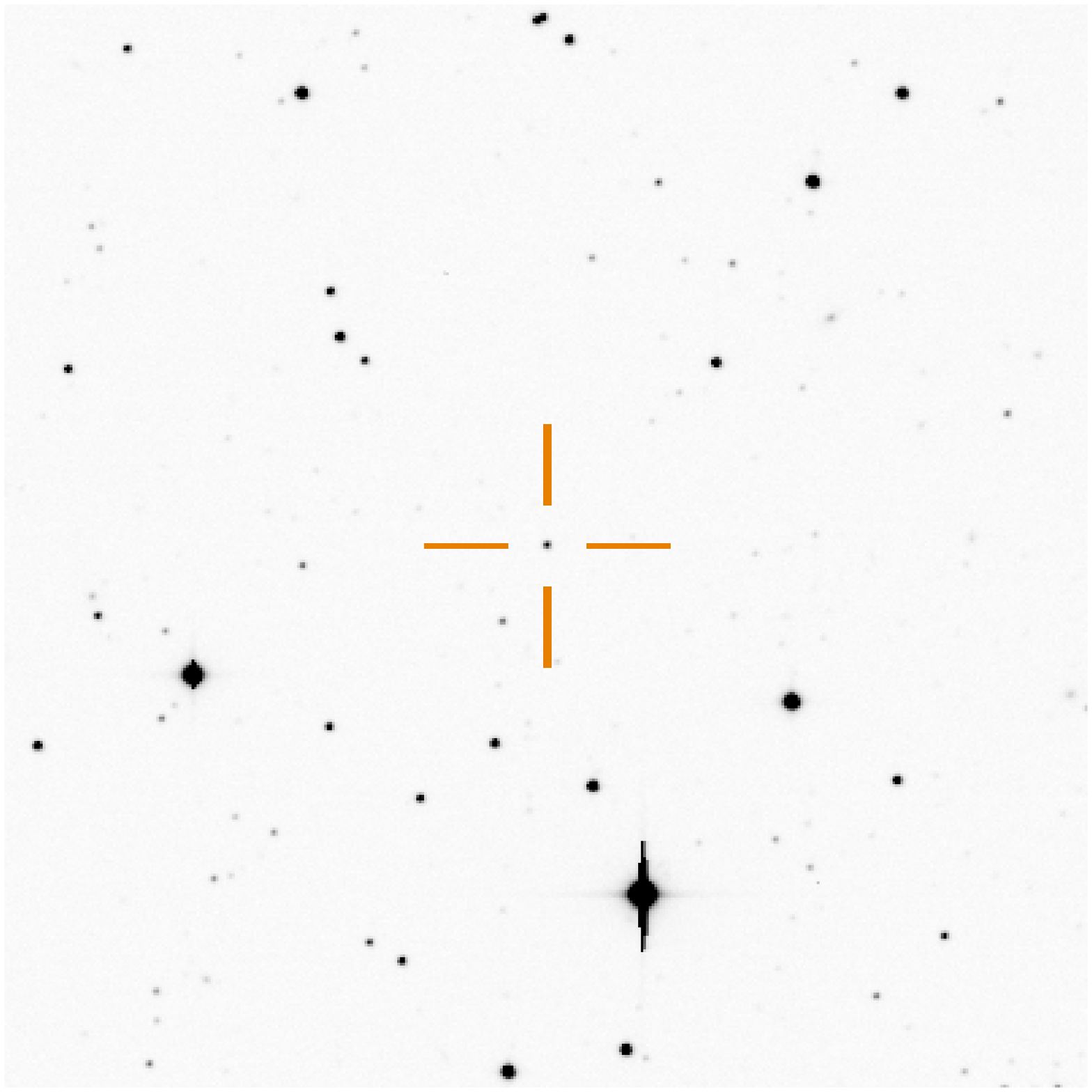}}}}\hspace{0.15in}
   \caption{Images of the newly-discovered systems taken with the PTF camera (\textit{left-to-right:} PTFEB11.441, PTFEB28.235 and PTFEB28.852). North is up and East is to the left; each image shows a 400$\times$400 arcsecond cutout from the 3.50 $\times$ 2.31 degree PTF camera field.}
   \label{fig:images}
\end{figure*}

\subsubsection{A Parallel Eclipse Search Using Graphics Processing Units}

Systems with smaller eclipse depths and/or longer periods and reduced
duty cycles may have only slightly increased photometric variability,
necessitating a more sensitive search. We use a standard
box-least-squares (BLS; \citet{Kovacs2002, Tingley2003}) algorithm to phase the light curves at
all possible periods and search for a transit-like signal. 

This algorithm requires the testing of thousands of periods and transit phases and is thus
computationally expensive. However, the problem is easily parallizable
as an arbitrary number of light curves and periods can be tested
simultaneously. We take advantage of this by implementing the BLS
algorithm on a Graphics Processing Unit that can perform hundreds of
computations in parallel.

Our BLS search is run on an NVIDIA Tesla C2050 Graphics Processing Unit that
contains 448 cores operating at 1.15 GHz, for a total of 1.03 Tflops in
single precision floating point arithmetic.  The BLS search is coded
in CUDA and is called from the PyCUDA python module \citep{Klockner2009}.

\begin{figure}
	\subfigure{\resizebox{\columnwidth}{!}{{\includegraphics{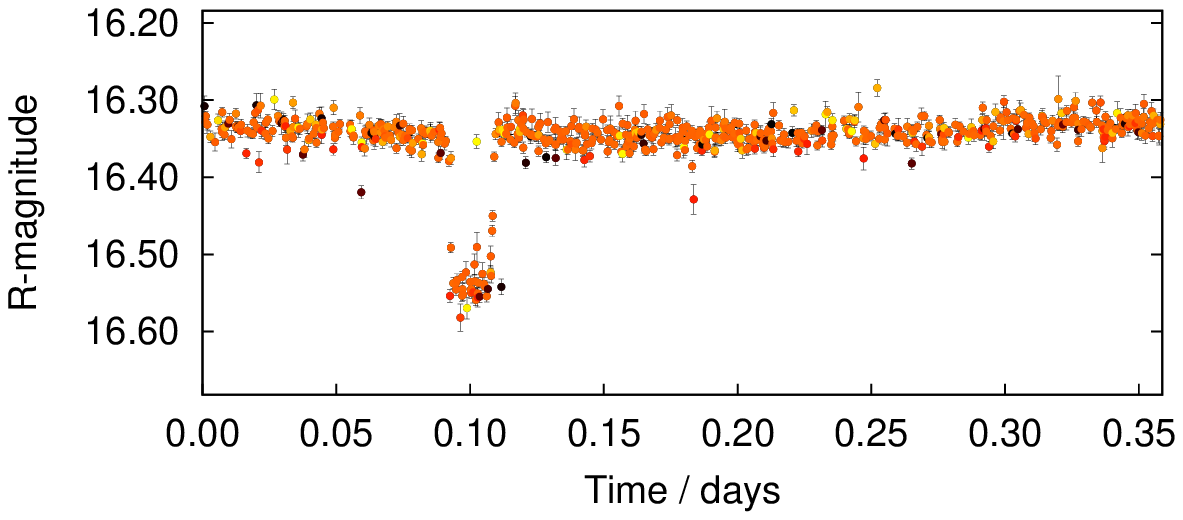}}}}\hspace{0.15in}
	\subfigure{\resizebox{\columnwidth}{!}{{\includegraphics{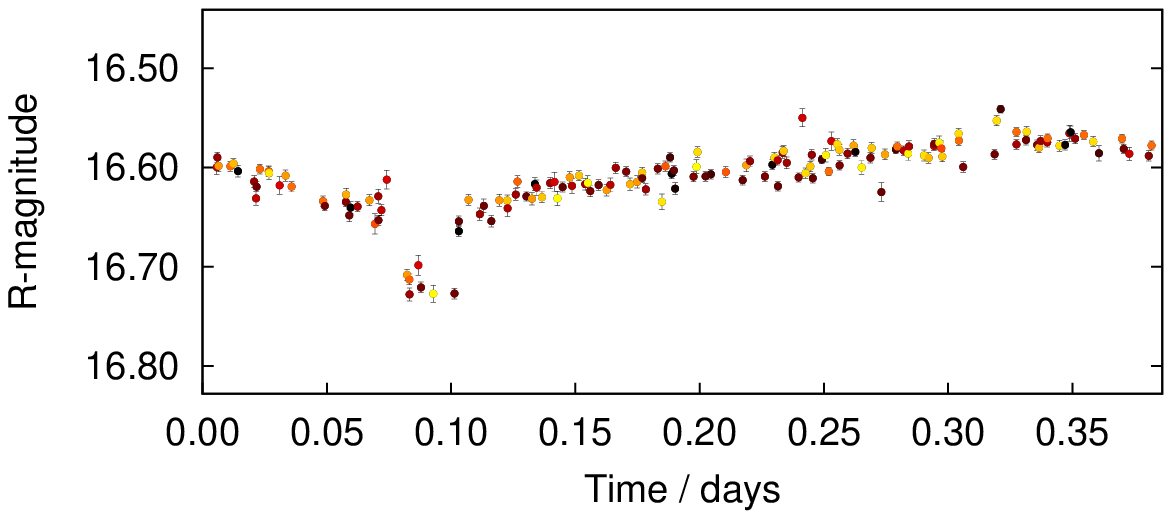}}}}\hspace{0.15in}
	\subfigure{\resizebox{\columnwidth}{!}{{\includegraphics{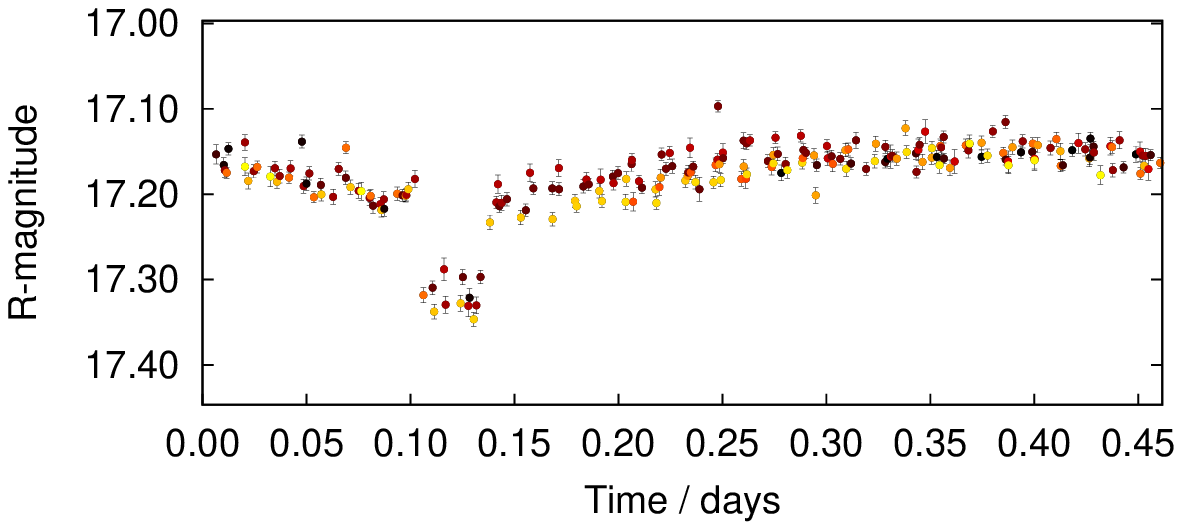}}}}\hspace{0.15in}

   \caption{The discovery P48 light curves for the three eclipsing systems (top-to-bottom: PTFEB11.441, PTFEB28.235, PTFEB28.852). The colouration of the points corresponds to the time the datapoint was measured; black are the oldest points and bright yellow are the newest. PTFEB11.441 was observed during a high cadence run targeted at M31, hence the large density of points all taken on a single night.}
   \label{fig:p48_lcs}
\end{figure}

\begin{figure}

  \resizebox{1.0\columnwidth}{!}
   {
	\includegraphics{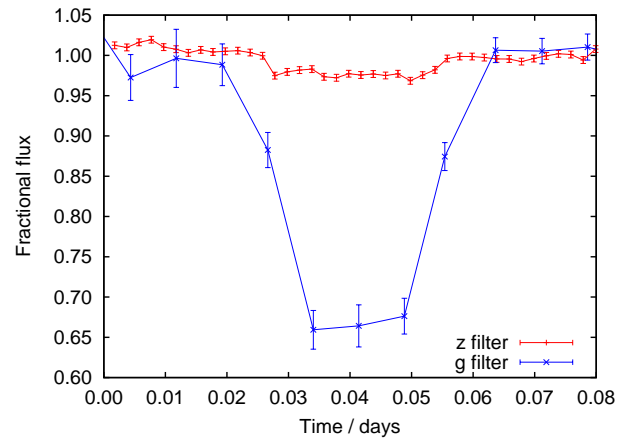}
   }
   \caption{Faulkes-North (z-filter) and BOS (g-filter) follow-up photometry of PTFEB28.852. The eclipse is detected at high significance in both bands, but is approximately 10$\times$ deeper in the g filter, suggesting the secondary eclipse of a hot body.}
   \label{fig:multicol}
\end{figure}

\begin{figure}

  \resizebox{1.0\columnwidth}{!}
   {
	\includegraphics{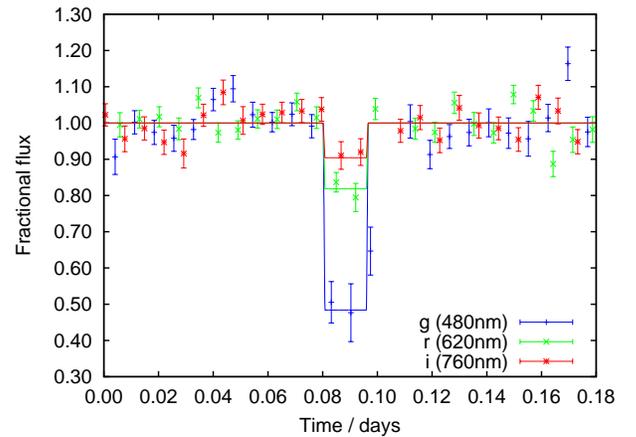}
   }
   \caption{Dunlap Institute Arctic Telescope photometry of PTFEB11.441, showing the varying eclipse depth with wavelength. The g, r and i filters (with midpoints shown in the figure) were alternated in sequence to produce a near-simultaneous multi-color light curve. To guide the eye, each color has a fit to the light curve using the system's eclipse parameters derived in section \ref{sec:high_speed}.}
   \label{fig:dit_multicol}
\end{figure}

Perhaps the simplest parallization technique for the BLS algorithm is to have each GPU
thread perform all the calculations for one source; in this way
several hundred sources could be tested simultaneously. However, the BLS algorithm requires
at least one access to the full light curve data at each test
period. The per-thread memory in typical GPUs is too small to contain a full light
curve, necessitating frequent calls to the GPU global memory. These
calls are slow, even when synchronized across threads, greatly limiting
the speed of the GPU implementation.

Instead, we consider sources sequentially, and simultaneously test
hundreds of different periods on a single source. The processing of each
light curve proceeds as follows: firstly, the light curve data is read into fast
shared memory by each thread block (with memory accesses coalesced for
speed), then each thread picks an untested period, phases the light
curve at that period and then tests a range of durations and
phases. 

In testing with typical PTF/M-dwarfs data, our GPU BLS algorithm
operates approximately eight times faster than a multi-threaded
program executing on all cores of an Intel Core-2 Quad Core CPU
running at 2.50 GHz. With the GPU approximately 10,000 light curves can be fully
searched for transit or eclipse events each hour.

\section{Discoveries and Follow up observations}
\label{sec:disc_and_followup}

We discovered three eclipsing M-dwarf / white dwarf binary systems in a search of $\approx$45,000 M-dwarfs from the first year of PTF/M-dwarf operations. Each system was detected as a high-confidence planet candidate, with a 5-20\% depth flat-bottomed eclipse in R-band with duration consistent with a transiting Jupiter-radius planet. Images of the field around each target are shown in figure \ref{fig:images}, and the P48 detection light curves of the systems are shown in figure \ref{fig:p48_lcs}.

The PTF names of the three detected sources are \mbox{PTF1 J004546.0+415030.0}, \mbox{PTF1 J015256.6+384413.4} and \mbox{PTF1 J015524.7+373153.8}. For brevity we hereafter refer to the sources by their PTF/M-dwarfs survey internal names, which are based on their decimal RA coordinates: \mbox{PTFEB11.441}, \mbox{PTFEB28.235} and \mbox{PTFEB28.852} respectively.

\subsection{Follow up measurements}
After the systems were detected in PTF data as planet candidates we followed the standard PTF/M-dwarfs follow up strategy: high-cadence multicolor photometry with the Palomar 60-inch, BOS, and FTN, along with radial velocity observations with the HIRES spectrograph on Keck-1. The multicolor photometry (section \ref{sec:follow_phot}) rapidly revealed that these objects had a strongly varying eclipse depth with color (figures \ref{fig:multicol} and \ref{fig:dit_multicol}), suggesting we were seeing eclipses between two self-luminous objects with different temperatures. Low-resolution spectra (section \ref{sec:follow_low_res_spec}) were sufficient to immediately confirm a white-dwarf component in one system (PTFEB11.441) but required detailed modelling to recover the white dwarf components in the other two systems (section \ref{sec:spec_model}). Radial velocity observations were scheduled to allow determinations of the white dwarf mass.

\subsubsection{High-cadence, multi-color photometry with BOS}
\label{sec:follow_phot}
    Multi-color photometry data for each of the targets were gathered with the RC Optics 0.8m telescope at Byrne Observatory at Sedgwick reserve near Santa Ynez, CA. The BOS telescope is equipped with a Santa Barbara Instrument Group STL-6303E camera utilizing a 3k x 2k Kodak Enhanced KAF-6306E CCD, with a 14.7' x 9.8' field of view and a pixel scale of 0.572" per pixel (2$\times$2 binning). We observed individual eclipses in a single color with SDSS i', SDSS g', or Astrodon Photometrics UV-blocked clear filters.

    The images were reduced using standard routines for bias subtraction, dark current subtraction, and flat-field correction. We extracted fluxes for all stars in the frame using aperture photometry routines in PyRAF. Relative flux light curves were produced by dividing the flux of the target star by the summed flux from several comparison stars in each image. Julian dates of mid-exposure were recorded during the observations, and later converted to barycentric Julian dates using the online tools described by \citet{Eastman2010}. Aperture sizes were optimized by minimizing the scatter of the resulting light curves, and ranged from 3" to 6". Exposure times were typically 300s.

Several eclipses of PTF11.441 were observed in short-cadence mode at BOS. We used an Astrodon Photometrics UV-blocked clear filter in order to maximize the signal to noise ratio (SNR) per unit time. We used 3$\times$3 binning (0.858" / pixel), and set the camera to only readout a small subsection of the CCD encompassing the target star, and three nearby stars of similar brightness to be used as comparison stars. This reduced the readout plus dead time of the instrument from $\approx$10s to $\approx$5s, and allowed us to achieve a photometric noise rate (PNR)\footnote{Photometric noise rate, calculated as RMS / $\sqrt(\Gamma)$, where RMS is the scatter of the out-of-eclipse section of the light curve and $\Gamma$ is the median number of cycles (exposure time and dead time) per minute. See \citet{Shporer2010b} for a more detailed description.} of $\rm \approx 1.6\% / min$. 

\subsubsection{Faulkes-North Photometry}
PTF28.852 was observed photometrically on 2010-11-10 and 2010-11-16 using the 2.0\,m Faulkes Telescope North (FTN, Maui, Hawai'i) operated by LCOGT. In both cases, the Spectral CCD imager\footnote{http://www.specinst.com} was used along with a Pan-STARRS-z filter. The Spectral instrument contains a Fairchild CCD486 back-illuminated $4096\times4096$ pixel CCD which was binned 2$\times$2 giving 0.303" pixels and a field of view of $10\arcmin \times 10\arcmin$. Exposure times were 130 \& 150\,seconds respectively on the two nights.

The frames were pre-processed using standard techniques for bias subtraction and flat-fielding; dark current subtraction was not performed as it is negligibly small for this instrument. Object detection and aperture photometry were performed using the DAOPHOT photometry package within the IRAF environment. The aperture sizes used were 7 and 5 pixels in radius on the two nights. Differential photometry was performed relative to 5--6 comparison stars within the field of view.

\subsubsection{Dunlap Institute Arctic Telescope Photometry}

The Dunlap Institute Arctic Telescope is a 20-inch robotic telescope currently undergoing testing at the New Mexico Skies observatory at Cloudcroft, NM. Once testing is complete, the telescope will be based at the PEARL research station on Ellesmere Island at a latitude of 80 degrees North, where it will perform a search for transiting planets. The telescope is equipped with a 16-megapixel Apogee U16M camera with a $34\arcmin \times 34\arcmin$ field of view.

We observed PTFEB11.441 on the night of October 23 2011. 120-second exposure g, r, and i-band images were taken in sequence throughout an eclipse of the target. After standard calibrations, differential photometry was performed using the pipeline described in section \ref{sec:p48_reduction}.

\subsubsection{Low-resolution Spectra}
\label{sec:follow_low_res_spec}
We obtained low-resolution optical spectra of PTFEB28.235 and PTFEB28.852 on 2010 Nov 30 UT, and PTFEB11.441 on 2011 Oct 25 UT. Spectra were acquired with the dual-arm Kast spectrograph \citep{ms93} on the 3-m Shane telescope
at Lick Observatory.  The spectra used a 2 arcsec wide slit, a 600/4310 grism on the blue side and a 300/7500 grating on the red side, yielding FWHM resolutions of $\approx 4$\,\AA\,and $\approx 10$\,\AA.  All observations were aligned along the parallactic angle to reduce differential light losses.

All spectra were reduced using standard techniques (e.g.,
\citealt{fps+03}).  Routine CCD processing and spectrum extraction
were completed within \texttt{IRAF}, and the data were extracted with
the optical algorithm of \citet{h86}.  We obtained the wavelength
scale from low-order polynomial fits to calibration-lamp spectra.
Small wavelength shifts were then applied to the data after
cross-correlating a template sky to the night-sky lines that were
extracted with the supernova.  Using our own \texttt{IDL} routines, we
fit spectrophotometric standard-star spectra to the data to flux
calibrate our spectra and to remove telluric lines
\citep{wh88,mfh+00}.

\begin{table}
\caption{High-Resolution Spectroscopic Observations}
\label{tab:rvs}
\centering

\begin{scriptsize}
\begin{tabular}{lcccrr}
Name & Epoch & Phase & $t_{int}$ & $v_{MD}$& EW(H$\alpha$)$_{MD}$\\
& (BJD$-$2400000) &  & (s) & (km s$^{-1}$) & ($\AA$)\\
\hline
PTFEB11.441&55739.535370&0.622&600&83.7&-0.99\\
PTFEB11.441&55739.562748&0.700&600&154.1&-0.62\\
PTFEB11.441&55739.601048&0.805&600&126.2&-0.75\\
PTFEB11.441&55740.534509&0.407&600&-120.6&-0.77\\
PTFEB28.235&55542.280580 &0.615 &600&121.0&-3.04\\
PTFEB28.235&55542.366785&0.838&600&154.1&-3.76\\
PTFEB28.235&55542.459073&0.077&600&-70.0&-3.24\\
PTFEB28.235&55543.252720&0.133&600&-132.7&-3.49\\
PTFEB28.235&55543.337034&0.351&600&-134.4&-3.39\\
PTFEB28.235&55740.592040&0.247&600&-209.1&-6.32\\
PTFEB28.852&55499.433465&0.288&900&132.0&-4.07\\
PTFEB28.852&55499.550803&0.539&900&-30.2&-3.51\\
PTFEB28.852&55499.588934&0.622&600&-101.8&-4.52\\
PTFEB28.852&55543.440246&0.638&600&-113.8&-3.17\\
PTFEB28.852&55500.400928&0.381&600&108.1&-3.93\\
PTFEB28.852&55542.272876&0.109&600&102.8&-4.47\\
\end{tabular}
\end{scriptsize}
\end{table}

\subsubsection{Keck/HIRES Radial Velocity Observations}
\label{rv_obs}
We obtained high-dispersion spectra for the whte-dwarf / M-dwarf systems using the
High-Resolution Echelle Spectrometer (HIRES) on the Keck-I 10m
telescope. HIRES is a single-slit echelle spectrograph permanently
mounted on the Nasmyth platform. All observations were performed using
the red channel, and span a wavelength range of 4300-8600 angstroms.
All were obtained using the C2 decker, which yields a spectral
resolution of $R \sim 45000$. We processed our HIRES data using the
standard pipeline MAKEE, which automatically extracts, flat-fields,
and wavelength-calibrates spectra taken in most standard HIRES
configurations. In table \ref{tab:rvs}, we list the epochs and exposure times for
all of our HIRES observations.

In order to remove small drifts in the wavelength calibration over the
course of the night, we cross-correlated each spectrum's telluric
features at 7600 angstroms with the telluric features for standard
stars selected from \citet{Nidever2002}, placing all observations into a common frame set by the Earth's
atmosphere. As we have previously shown \citep{Kraus2011}, this reduces the systematic uncertainty in the measured velocities to $\sim$0.3 km/s. We then measured the position and equivalent width (EW) of the H-alpha line at each epoch. The results are summarized in table \ref{tab:rvs}.

\begin{figure*}
  \centering
	\subfigure{\resizebox{0.3\textwidth}{!}{{\fbox{\includegraphics{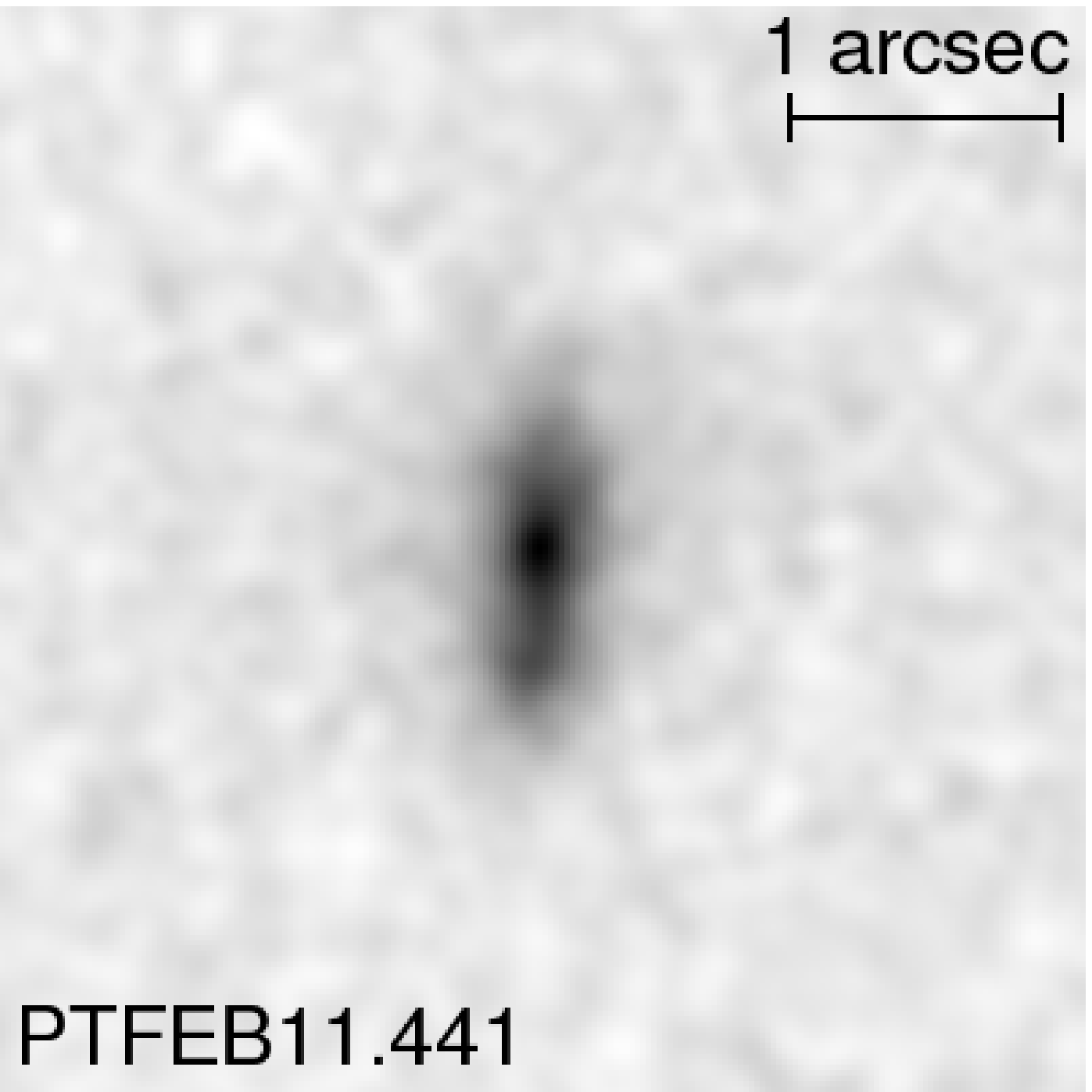}}}}}\hspace{0.05in}
	\subfigure{\resizebox{0.3\textwidth}{!}{{\fbox{\includegraphics{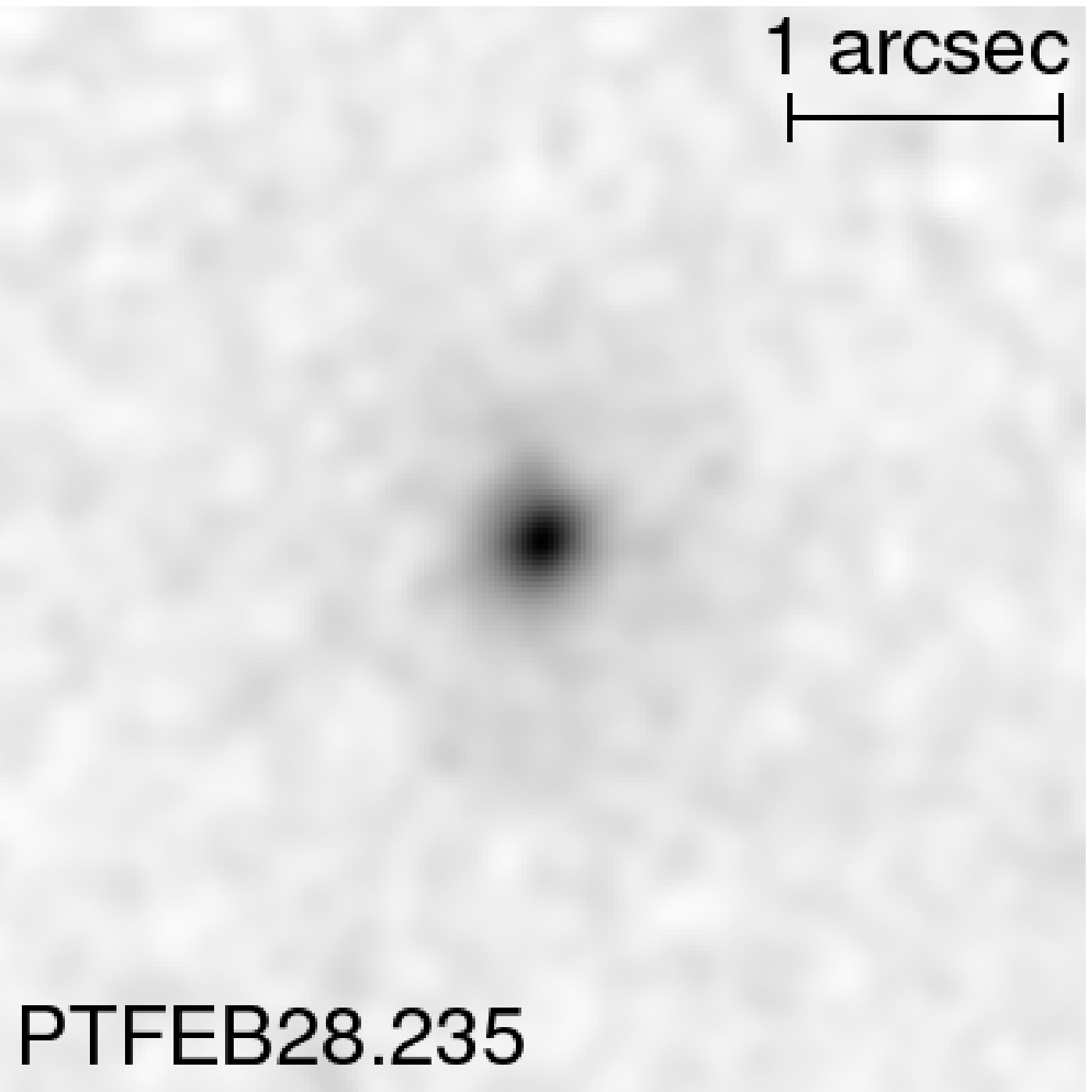}}}}}\hspace{0.05in}
	\subfigure{\resizebox{0.3\textwidth}{!}{{\fbox{\includegraphics{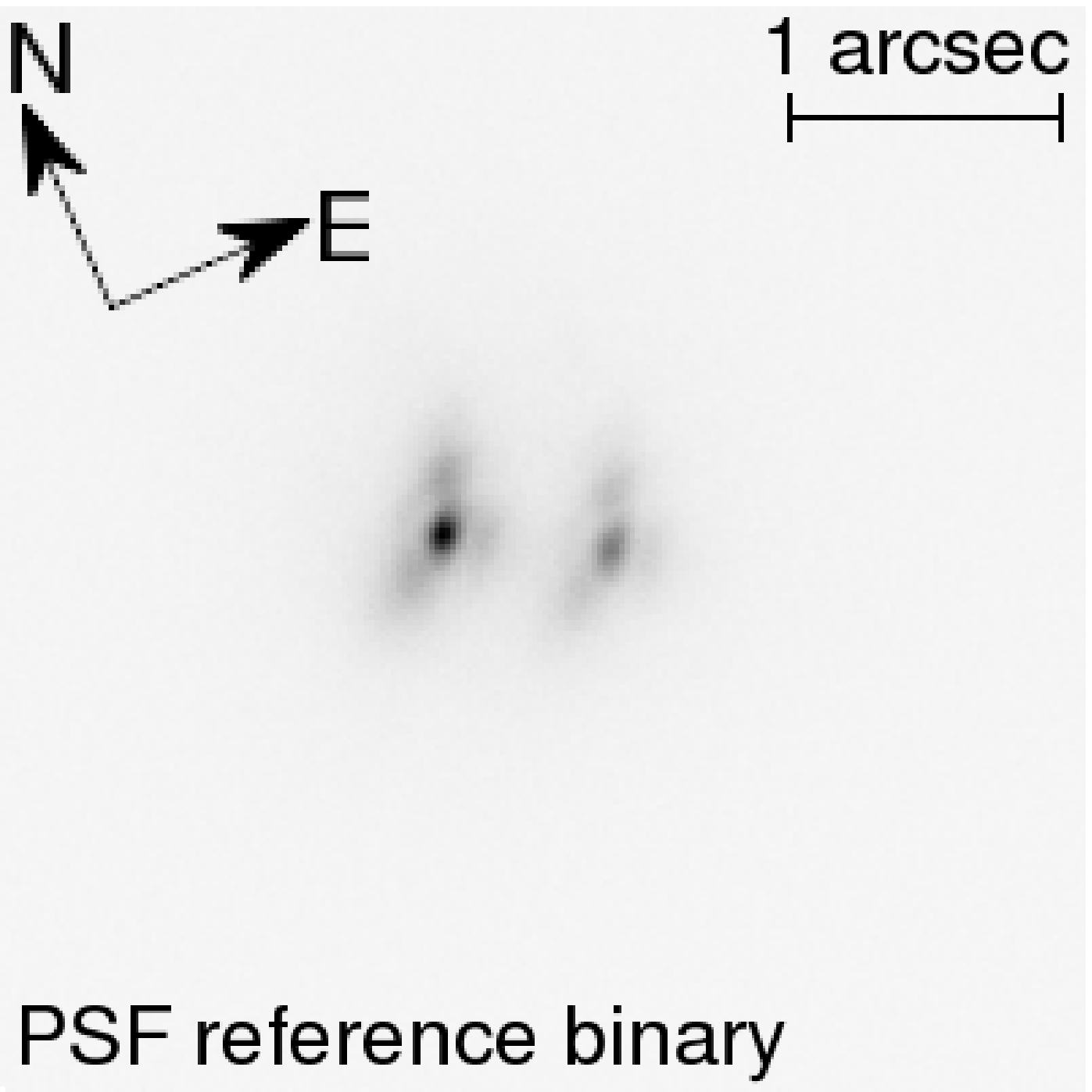}}}}}\hspace{0.05in}

   \caption{Robo-AO laser guide star adaptive optics images of (left-to-right) PTFEB11.441, PTFEB28.235, and an $\rm m_r\sim13$ binary taken in the same part of the night to provide a comparison point spread function. All images are in a 600nm long-pass filter, are 4 arcseconds square, have the same rotation, and are displayed with linear scaling. To improve the signal-to-noise for these faint targets, the PTFEB11.441 and PTFEB28.235 images have been smoothed by a Gaussian filter with a FWHM corresponding to the diffraction-limited resolution of the telescope. The apparent vertical smearing of the PTFEB11.441 image is due to the shift-and-add algorithm occasionally locking on to the secondary component of the binary \citep{Law2007}.}
   \label{fig:roboao}
\end{figure*}

\subsubsection{Robo-AO Laser Guide Star Adaptive Optics Observations}
Robo-AO is a visible and near-infrared laser guide star adaptive optics system specifically engineered for 1-3 m class telescopes \citep{Baranec2011}. The Robo-AO system comprises an ultraviolet Rayleigh laser guide star, an integrated adaptive optics and science camera system, and a robotic control system. The system currently incorporates both an electron-multiplying CCD (EMCCD) and an InGaAs infrared array camera for imaging. 

Robo-AO obtained images of PTFEB11.441 and PTFEB28.235 during a Robo-AO commissioning run at the Palomar 60-inch telescope on the night of 2011 Nov 18 UT. Robo-AO was operated without tip-tilt correction, instead relying on post-facto shift-and-add processing of the individual frames. We used a long-pass filter with a 600nm cut-on to obtain increased signal compared to a bandpass filter, along with relatively long 1-second exposure times to compensate for the faintness of the targets ($\rm m_R > 16$). We used the Lucky Imaging plus adaptive optics pipeline described in \citet{Law2009_ao} to perform the image alignment, along with frame selection at the 10\% level to boost the imaging resolution. After processing, images with approximately 0.2'' full-width half-maximum (FWHM) point spread functions were obtained on these targets during $\sim$1'' seeing conditions. These observations were limited by residual tip-tilt errors during the 1-second exposures; similar bright targets observed the same night with 33-ms exposures had a diffraction-limited angular resolution of $\approx$0.13 arcsec FWHM (figure \ref{fig:roboao}). 

The Robo-AO imaging reveals a bright companion to PTFEB11.441, located at 0.43$\pm$0.06 arcsec and at a position angle of 165$\pm$3 degrees East of North.

\begin{table*}
\caption{Measured and Estimated System Properties}
\label{tab:targets_hr}
\centering

\begin{scriptsize}
\begin{tabular}{lllllllllll}

\hline
Name & {$\rm RA$} & {$\rm Dec$} & {Period} & $\rm T_0$ & {$\rm SpT_{MD}$} & {$\rm Phot. SpT_{MD}$} & {$\rm T_{WD}$} & {Dist./pc} & {$\rm M_{MD}$} & {$\rm R_{MD}$} \\
     &            &             &          & (HJD)     & $(\pm 0.5)$ & $ (\pm 0.5)$ & $(\pm 500K)$ & & &\\
\hline
PTFEB11.441 & 00 45 46.0 & +41 50 30.0  &  0.35871d$\pm$0.00005d & 2455438.3165 & M3    & M3.5 & 8500K             & 180$\pm45^{*}$ & 0.35$\pm$0.05 $\rm{M_{\odot}}$ & 0.33$\pm$0.05 $\rm{R_{\odot}}$\\
PTFEB28.235 & 01 52 56.6 & +38 44 13.4  & 0.38611d$\pm$0.00011d & 2455460.1116 & M3    & M3 & 8000K & 200$\pm$50 & 0.35$\pm$0.05 $\rm{M_{\odot}}$ & 0.33$\pm$0.05 $\rm{R_{\odot}}$\\
PTFEB28.852 & 01 55 24.7 & +37 31 53.8  & 0.46152d$\pm$0.00009d & 2455530.7335 & M3    & M2 & 8500K & 260$\pm$70 & 0.35$\pm$0.05 $\rm{M_{\odot}}$ & 0.33$\pm$0.05 $\rm{R_{\odot}}$\\

\end{tabular}
\end{scriptsize}
\tablecomments{The properties of the detected systems, based on photometry and low-resolution spectroscopy. $\rm SpT_{MD}$ is the M-dwarf spectral type derived from low-resolution spectra, and $\rm Phot. SpT_{MD}$ is the spectral type estimated from infrared colors. The white dwarf properties are estimated as described in the text. The distance estimate for PTFEB11.441 is starred because it is likely to be an underestimate because of its possible companion, which is unresolved in the 2MASS photometry used to generate the distance estimate.}
\label{tab:targets}
\end{table*}

\subsection{The nature of the detected objects}

Multi-color through-eclipse photometry of the three detected systems showed that the eclipse depths were much larger in blue bands than in red bands. The sharp ingress and egress of the eclipses suggest a small body, while the flat-bottomed shape argues against that body eclipsing the M-dwarf, as limb-darkening would then be apparent.  Radial velocity observations showed a very large amplitude of more than 100km/sec. These observations suggest the presense of a small, hot, massive object undergoing total eclipses in these systems. The white-dwarf / M-dwarf binary scenario is consistent with all these observations.

The properties of the systems are summarized in table \ref{tab:targets}. We detail each system below.

\subsubsection{PTFEB11.441}

This system shows a possible companion in Robo-AO images. If it is associated, the companion is quite wide, at approximately 80 AU separation. There are no other sources detected in the Robo-AO image, which covers 1600$\rm arcsec^2$. On this basis, the probability of an unassociated star appearing within 2 arcsec of our target is less than 1\%. Definitive confirmation will require colors and common-proper-motion measurements, but for the purposes of the following sections we tentatively associate the companion star with PTFEB11.441. 

The two wide components of the system have roughly equal brightness and the low-resolution spectrum of the system is well fit by a combination of a white-dwarf and an M-dwarf spectrum without further components (section \ref{sec:spec_model}), suggesting that spectral types of the two M-dwarfs are very similar. Wide M-dwarf binaries make up a small fraction of the binary population, but many examples have been found \citep{Dhital2010}, and a very large fraction of the components of the widest binaries have close companions with roughly equal masses \citep{Law2010_wide}. However, we note that the M-dwarf near the white dwarf in PTFEB11.441 survived the common-envelope phase of the system's evolution, and so it is unlikely that this system formed in the same manner as other known wide M-dwarf binaries.

PTFEB11.441 lies in a PTF field targeted at M31 and is located approximately one degree from the center of the galaxy. Of the three systems, PTFEB11.441 shows the greatest amount of emission from the white dwarf, with the M-dwarf being overwhelmed at wavelengths $\lsim550nm$. The light curve shows only very-low-level out-of-eclipse variability. The target is detected by GALEX with a near-UV magnitude of $\rm {m_{NUV}=17.35}$, with no far-UV detection.

\subsubsection{PTFEB28.235}

The P48 light curve of this system shows out-of-eclipse variability at the 7-10\% level, almost as strong as the eclipse itself in R-band. The variability phasing (maximal when the white dwarf is in front of the M-dwarf) suggests that the variability is primarily an irradiation effect. The 5\% eclipse depth is the smallest R-band depth of the three systems. The system has a GALEX Near-UV detection at $\rm {m_{NUV}=21.6\pm0.4}$ (with no far-UV coverage).

\subsubsection{PTFEB28.852}

This system is very similar to PTFEB28.235, with similar variability levels and a similar light curve shape. The system has a GALEX far-UV detection at $\rm{m_{FUV}=21.0\pm0.1}$ (with no near-UV coverage).

\section{Mass and Radius Models and Measurements}
\label{sec:models}
The optical light in these systems is dominated by the M-dwarfs, and the large radius ratios between the components produce very small (millimag-level) primary eclipses. For these reasons, we were able to measure precise eclipse shapes only for the white-dwarf occultation events, and radial velocities for the M-dwarf components. However, the low-resolution spectra allow us to estimate the mass and radius of the M-dwarf component of the system and thus estimate the properties of the white dwarf component. Finally, high-cadence photometry of the ingress and egress of the white-dwarf occultation events allowed us to set stringent upper limits on the radius of the white dwarf in PTFEB11.441.

\subsection{Primary and secondary spectra}
\label{sec:spec_model}
We model the low-resolution spectra of our targets as a combination of a M-dwarf and a white dwarf spectrum. Template M-dwarf spectra are taken from the HILIB stellar flux library \citep{Pickles1998} and have steps of a single spectral subclass. The white dwarf models\footnote{\citet{Koester2010}. Balmer lines in the models were calculated with the modified Stark broadening profiles of \citet{Temblay2009}, kindly made available by the authors.} were kindly provided by Detlev Koester and are on a grid with 250K steps between 6000K and 10000K (and longer steps up to 20000K), and 0.25 dex steps between log g = 6.0 and log g = 9.5.

\begin{figure*}
  \centering
	\subfigure{\resizebox{0.46\textwidth}{!}{{\includegraphics{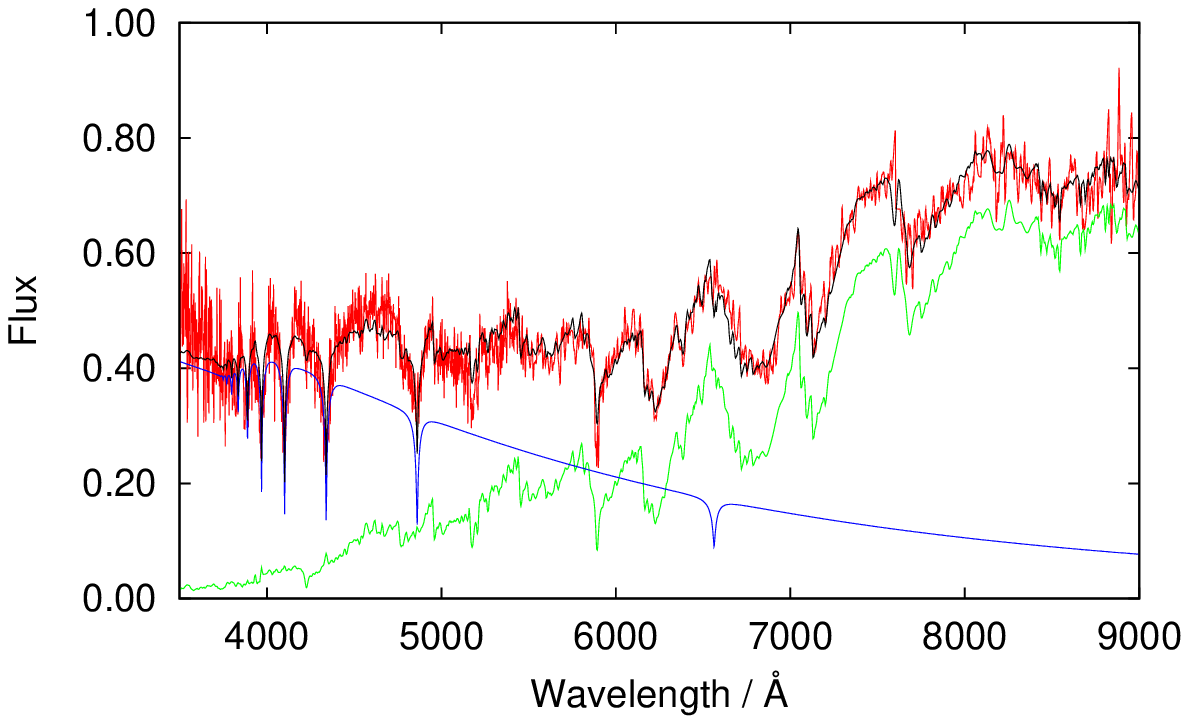}}}}\hspace{0.15in}
	\subfigure{\resizebox{0.46\textwidth}{!}{{\includegraphics{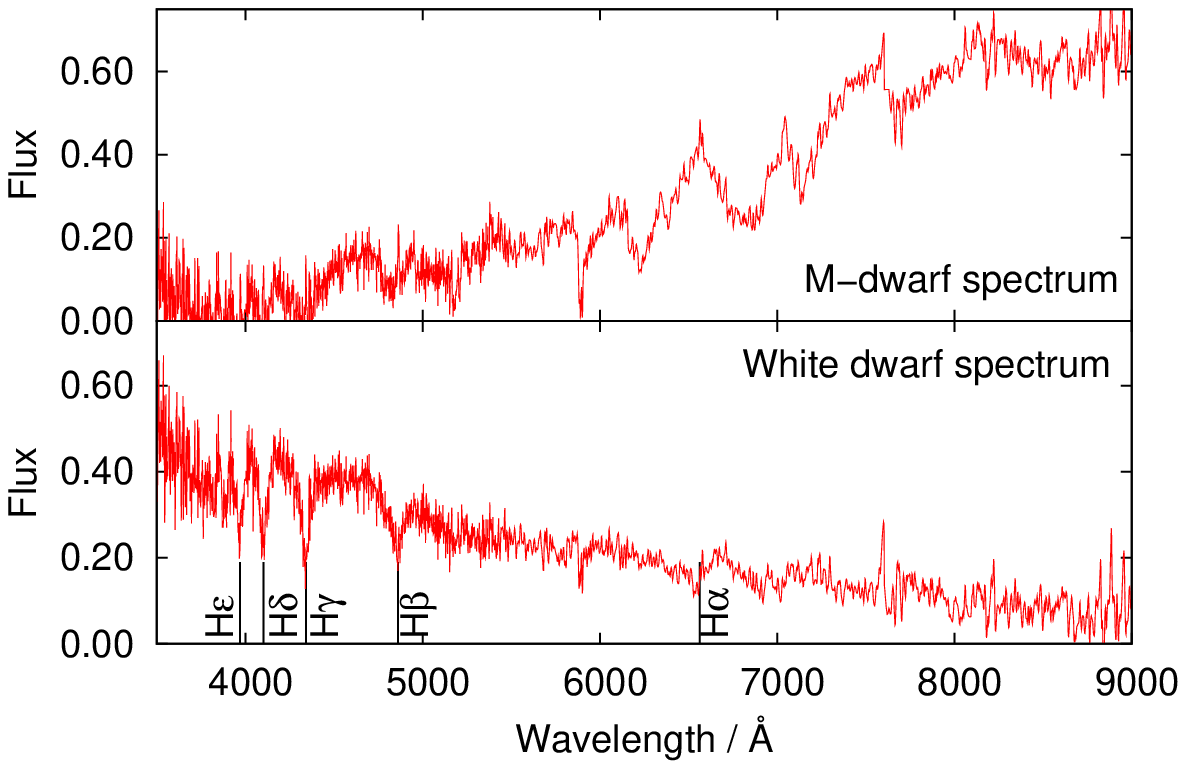}}}}\hspace{0.15in}

	\subfigure{\resizebox{0.46\textwidth}{!}{{\includegraphics{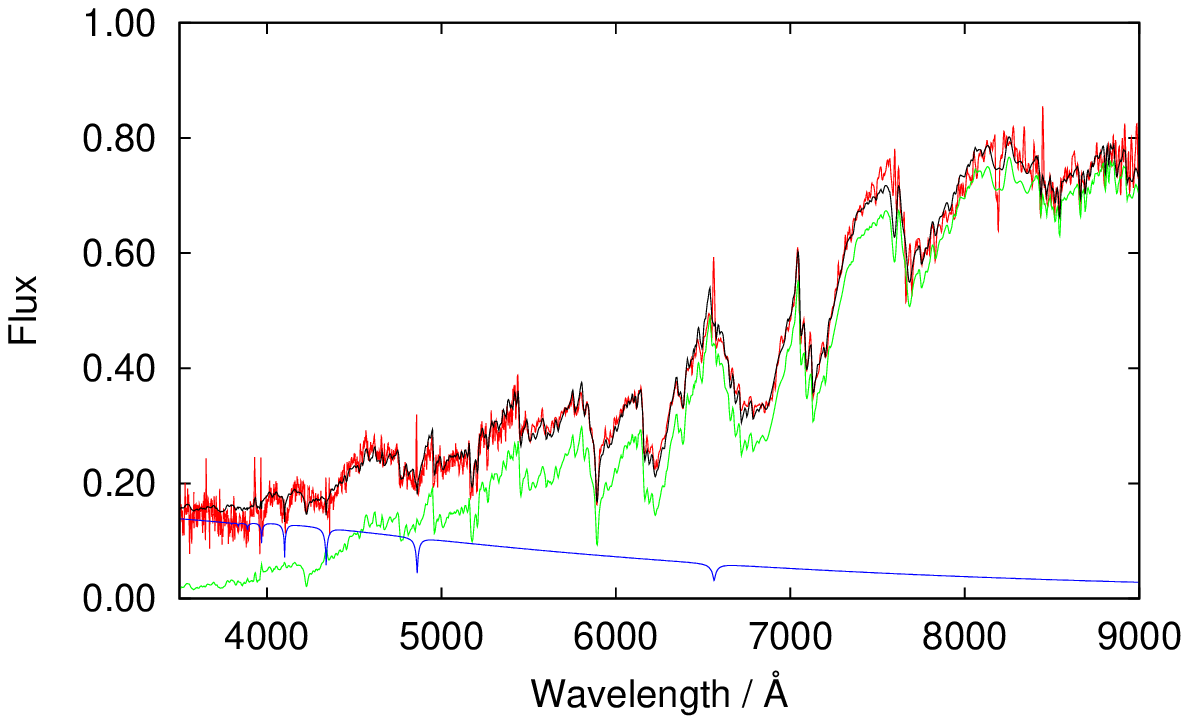}}}}\hspace{0.15in}
	\subfigure{\resizebox{0.46\textwidth}{!}{{\includegraphics{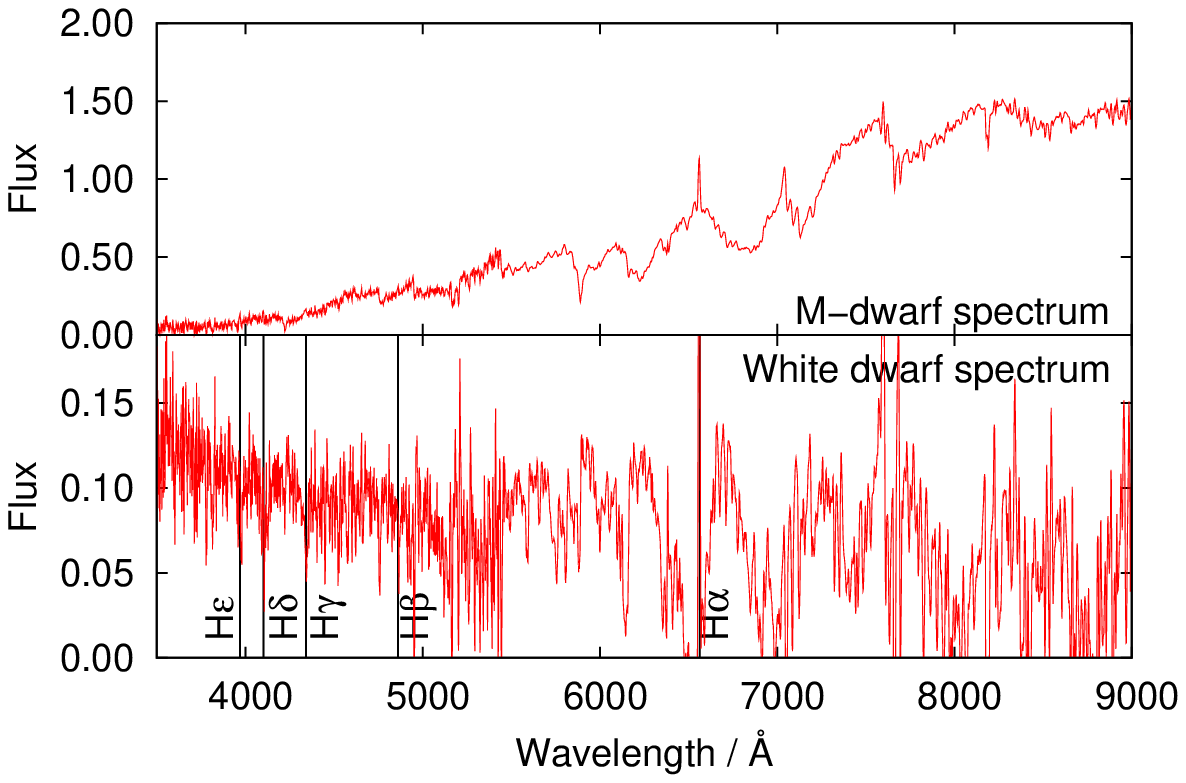}}}}\hspace{0.15in}

	\subfigure{\resizebox{0.46\textwidth}{!}{{\includegraphics{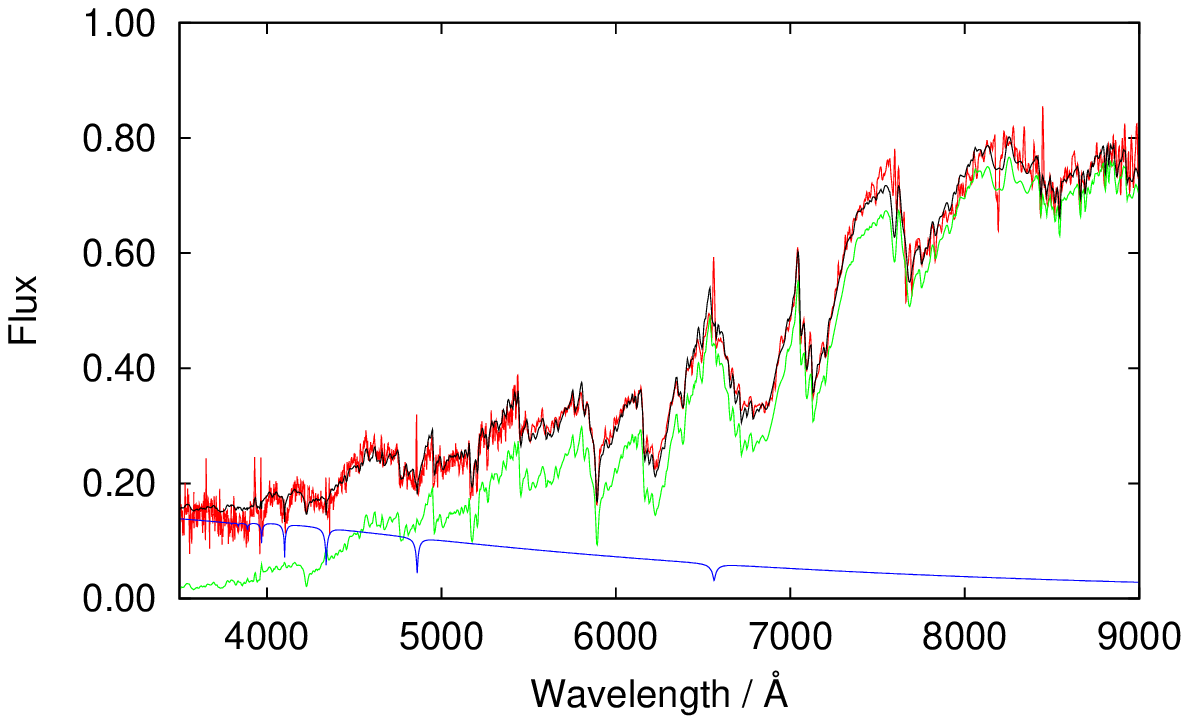}}}}\hspace{0.15in}
	\subfigure{\resizebox{0.46\textwidth}{!}{{\includegraphics{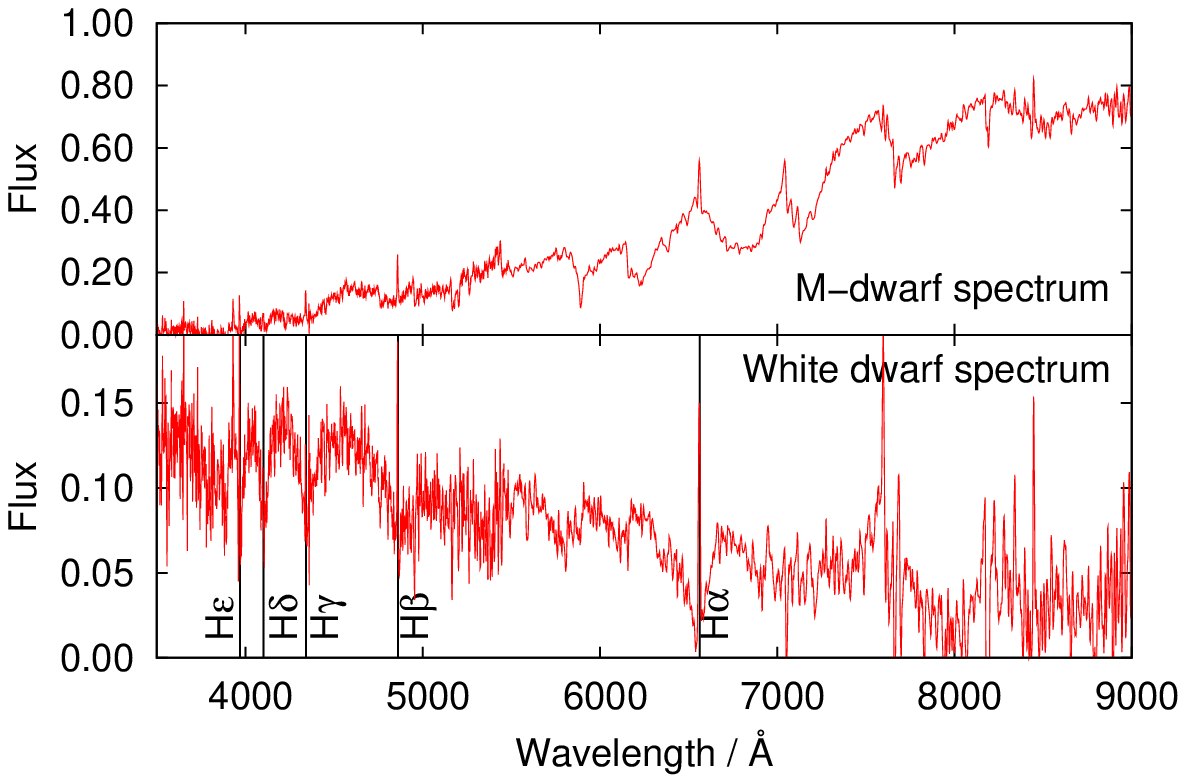}}}}\hspace{0.15in}

   \caption{Spectra for the three targets: in order from top to bottom, PTFEB11.441, PTFEB28.235, PTFEB28.852. \textit{Left:} the observed spectrum (red); the M-dwarf (green) and white-dwarf (blue) spectra simultaneously fit to the spectrum, and the fitted combined spectrum (black). \textit{Right:} separated spectra -- the observed spectrum with each model spectrum removed in turn. The Balmer lines in the white dwarf spectra are labelled.}
   \label{fig:specs}
\end{figure*}

We simultaneously fit the white dwarf and M-dwarf spectra using a downhill simplex algorithm and a bootstrap (e.g. \citealt{Press1992}) method to find the uncertainties in the fit. Figure \ref{fig:specs} shows the fitted models and each component of the binaries with the model for the other component subtracted. In each case subtraction of the template M-dwarf spectrum revealed a (high-noise) DA white dwarf spectrum with clear Balmer lines. 

PTFEB11.441 is well-fit by a simple combined white-dwarf and M-dwarf spectrum, suggesting that the third component in the system is also an M-dwarf, and that the two M-dwarfs have similar spectral types. In the following sections, we use the measured spectral type for the M-dwarfs in PTFEB11.441, assuming that it is the spectral type of the M-dwarf closest to the white dwarf.

\subsection{Reddening}
The M-dwarf spectral types measured from the two-component fit to the system spectra closely agree with those estimated from infrared photometry of the systems (table \ref{tab:targets}). This suggests that reddening is not a significant problem for the M-dwarf spectral types and thus mass and radius estimations. 

However, we note that reddening will affect the estimated white dwarf temperature more severely as its flux is mostly in the blue end of the spectrum. The estimated galactic extinction along the line of sight to PTFEB11.441 is E(B-V)=0.02 \citep{Schlegel1998}, and will be lower for the target itself which is located at a distance of $\approx400pc$. The other two targets are located in regions of higher line-of-sight extinction, at E(B-V)=1.5 and 1.1 mag for PTFEB28.235 and PTFEB28.852 respectively. The actual extinction for the two targets is, however, likely to be much lower as they are located at distances of only $\approx$200pc. This is supported by several different arguments: the very good match of the M-dwarf model spectrum to the observed spectrum; the match between the infrared-color-estimated spectral types and the spectrally-measured spectral types; and the GALEX detections of both objects, which require a small near and far-UV extinction.

\subsection{M-dwarf Properties}
We fit the M-dwarf masses, radii and distances in the following manner: we first fit survey photometry of the system with the stellar SEDs listed in \citet{Kraus2007a}, yielding a photometric spectral type and distance. Despite the availability of other photometric data such as USNO-B1, we restrict our fits to 2MASS J, H and K colors that are unlikely to be affected by the light from the white dwarf. We assume a 25\% distance uncertainty \citep{Kraus2007a, Law2010}, and $\pm$1-2 spectral subclass uncertainty \citep{Kraus2007a}; the SED estimates for the spectral types are consistent with those derived from the two-component fits to our low-resolution spectroscopy. We also attempted to determine the M-dwarf spectral type from the TiO5 narrowband spectroscopic index \citep{Gizis1997}, but found implausibly high values of the index, suggesting possible contamination by white dwarf emission. In the following sections we adopt the spectral types estimated from the two-component fits to the low-resolution spectra, as that method yields the lowest spectral type uncertainty. 

We estimate the M-dwarf masses from the spectral type vs. mass calibrations detailed in \citet{Delfosse2000}, assuming they are on the main-sequence and have solar metallicity and age. All three of our targets have a measured spectral type of M3, giving an estimated mass of 0.35$\pm$0.05 $M_{\odot}$. We note that a small fraction of the systems measured in \citet{Delfosse2000} have much lower masses at the M3 spectral type, for reasons that are still unclear. To estimate the masses in an alternate manner, we estimated the M-dwarf effective temperatures from the relation described in \citet{Luhman1999}. We then combined those estimates with the 5GYr isochrones of solar-metallicity stars in \citet{Baraffe1998} to estimate the stellar masses. These relations also predict a much lower mass for our M-dwarf targets, of 0.23$^{+0.06}_{0.04} \rm{M_{\odot}}$. In the following sections we adopt the higher masses measured in \citet{Delfosse2000}. We, however, note that two-component RV measurements for our systems giving masses for each component would help understand the mass / spectral type relation in this mass range.

We estimate the M-dwarfs' radii from a fit to the eclipsing-binary-derived mass/radius relation. We note that M-dwarfs in close binaries appear to be biased towards larger radii compared to systems with wider orbital radii \citep{Kraus2011}. We adopt a mass/radius relation for close ($<$ 1 day period) binaries based on the measurements summarized in \citet{Kraus2011}:
\begin{equation}
\rm R = 1.096M - 0.052 \mbox{\,\,\,\,\,\,\,for\,\,\,\,\,\,\,} 0.1 M_{\odot} < M < 0.7 M_{\odot} 
\end{equation}
where M is the mass in solar masses and R is the radius in solar radii. The radii in this relation are approximately 5-10\% larger than both those measured for larger orbital radius systems and theoretical expectations (e.g. \citet{Baraffe1998, Kraus2011}). The results are summarized in table \ref{tab:targets}.

\begin{figure*}
  \centering
	\subfigure{\resizebox{0.31\textwidth}{!}{{\includegraphics{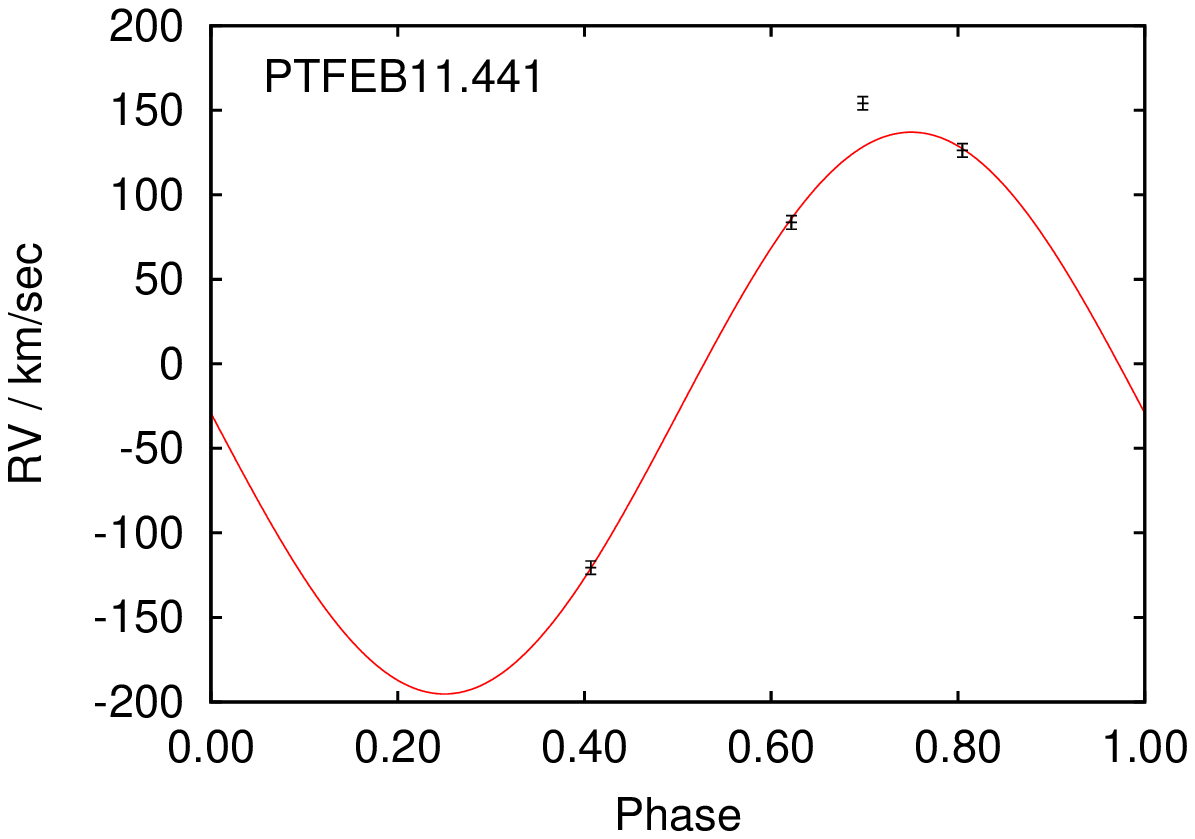}}}}\hspace{0.15in}
	\subfigure{\resizebox{0.31\textwidth}{!}{{\includegraphics{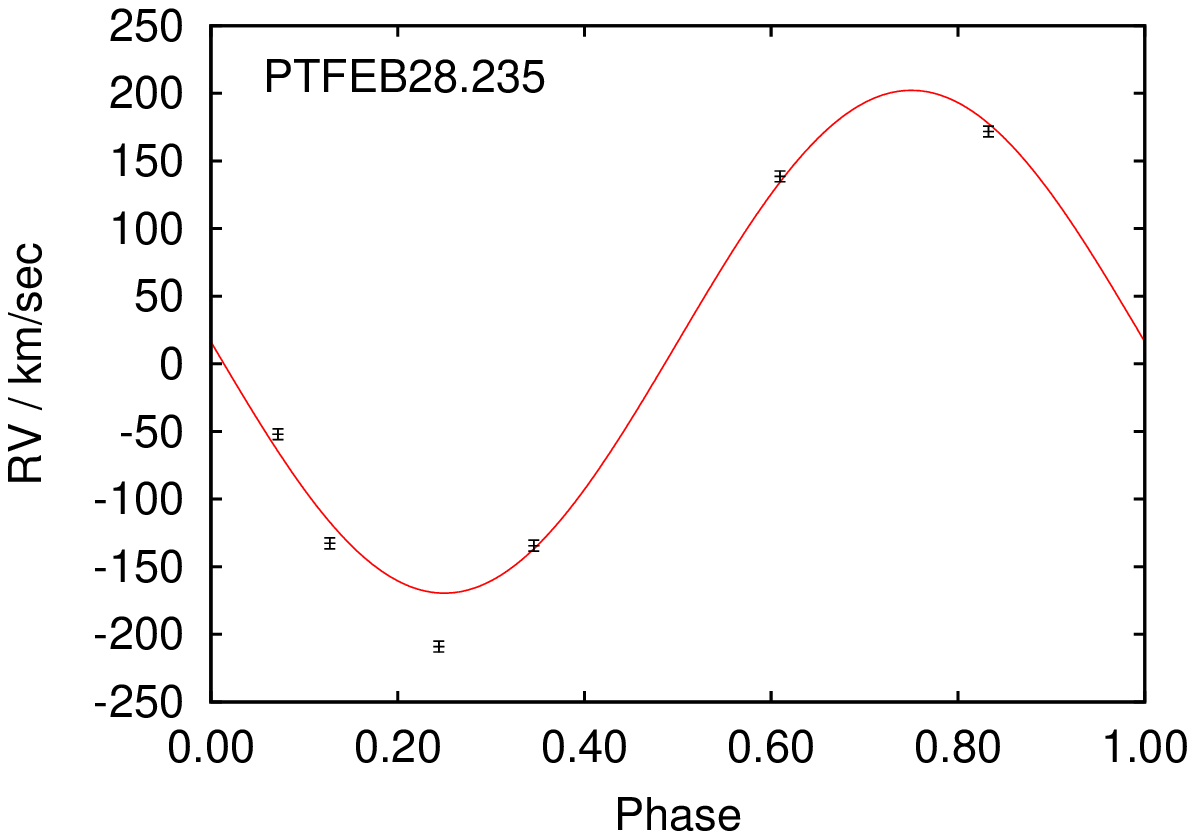}}}}\hspace{0.15in}
	\subfigure{\resizebox{0.31\textwidth}{!}{{\includegraphics{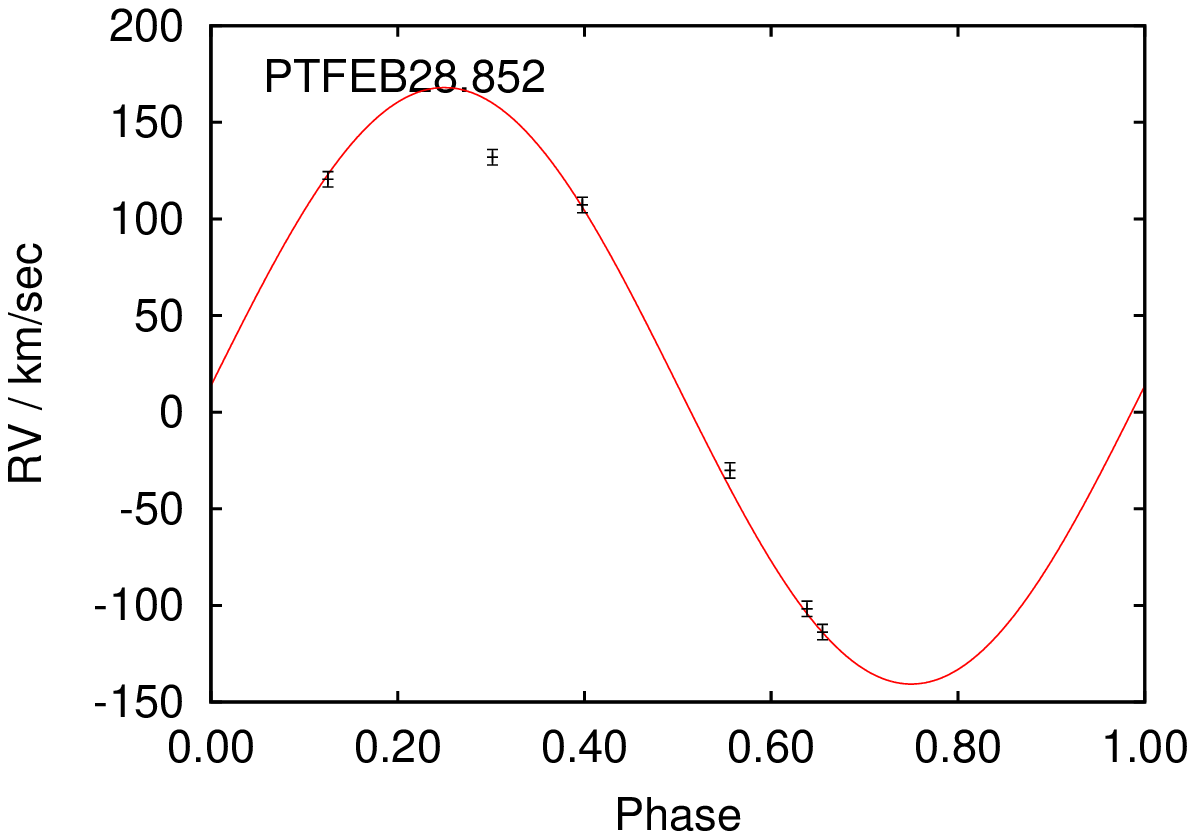}}}}\hspace{0.15in}

   \caption{Radial velocity orbit fits for each of the targets. Each binary shows an anomalous measurement around the quadrature points, which may be due to the white-dwarf's H$\alpha$ absorption line interfering with the measurement of the M-dwarf H$\alpha$ position. A simple radial velocity orbit is consistent (reduced $\chi^2 \approx 1$) with all the other points, but inconsistent with the anomalous points, and so we remove those points to avoid unduly biasing the fits.}
   \label{fig:rvs}
\end{figure*}

\begin{table}
\caption{Spectroscopic orbit properties}
\label{tab:rv_orbits}
\centering
\begin{scriptsize}
\begin{tabular}{llll}
\hline
{\bf Name} & {\bf $\rm K_1$ (km/sec)} & {\bf $\rm V_0$ (km/sec)} & {\bf Mass function (F)}\\
PTFEB11.441 & -166$\pm$15    & -29$\pm$26 & 0.14$\pm$0.04\\
PTFEB28.235 & -186$\pm$17    & 16$\pm$15 & 0.26$\pm$0.07\\
PTFEB28.852 & 154$\pm$5    & 13$\pm$3 & 0.17$\pm$0.02\\
\end{tabular}
\end{scriptsize}
\label{tab:spec_orbits}
\end{table}

\subsection{White Dwarf Properties}
We measure the white dwarf masses from the measured radial velocity curves, assuming the M-dwarf masses derived above.  We fit a simple model to the velocity curves (figure \ref{fig:rvs} and table \ref{tab:spec_orbits}), assuming zero eccentricity because of the short circularization timescales for these compact systems (for example, see \citealt{Devor2008}). PTFEB28.235 and PTFEB28.852 have sufficient numbers of RV points to allow an eccentric orbit fit, but the fitted eccentricities were consistent with being zero. The white dwarf masses ($M_{WD}$) are related to the fit mass function (F), the M-dwarf masses ($M_{MD}$) and the inclination (i) by:
\begin{equation}
{\rm F} = \frac{M_{WD}^3 \sin^3{i}}{(M_{MD} + M_{WD})^2} 
\end{equation}
which can be solved as a cubic equation. The results are summarized in table \ref{tab:wd_mass_radii}. The inclination has only a weak effect on the derived white dwarf masses; we marginalize it across all ranges allowed by the eclipse geometry.

We estimate the white dwarf radii in two ways: firstly, models of the white dwarf effective temperature, mass and surface gravity cooling curves, and secondly a direct fit to the white dwarf's eclipse light curve.

Our model radius estimates are based on the Bergeron cooling curves for DA white dwarfs \citep{Holberg2006, Kowalski2006, Tremblay2011, Bergeron2011}, for which we derive the allowed surface gravity range from the white dwarf mass range measured above. The effective temperature has only a small effect on the surface gravity estimates and so we fix it at the best-fit value from the low-resolution spectra.

\begin{table}
\caption{White dwarf properties}
\centering
\begin{scriptsize}
\begin{tabular}{lllll}
\hline
{\bf Name} & ${\bf M_{WD}}$ & ${\bf R_{WD, model}}$ & ${\bf R_{WD, spec.}}$ & ${\bf R_{WD, direct}}$\\
PTFEB11.441 & 0.51$\pm$0.09$\rm{M_{\odot}}$ & 0.014$\pm$0.002$\rm{R_{\odot}}$  & 0.010$\rm{R_{\odot}}$ & $<$0.025$\rm{R_{\odot}}$ (95\%)\\
PTFEB28.235 & 0.65$\pm$0.11$\rm{M_{\odot}}$ & 0.012$\pm$0.002$\rm{R_{\odot}}$  &\nodata & \nodata\\
PTFEB28.852 & 0.52$\pm$0.05$\rm{M_{\odot}}$ & 0.014$\pm$0.001$\rm{R_{\odot}}$  &\nodata & \nodata\\
\end{tabular}
\end{scriptsize}
\tablecomments{Derived masses and radii for the white dwarfs in the eclipsing binary systems. ${\bf M_{WD}}$ is the stellar mass; ${\bf R_{WD, model}}$ is the radius derived from the mass and the Bergeron cooling models; ${\bf R_{WD, spec.}}$ is the radius derived from direct fits of the \citet{Koester2010} model spectra Balmer lines to the white dwarf spectrum; ${\bf R_{WD, direct}}$ is the 95\%-confidence upper limit measured from the eclipse ingress and egress shapes.\\
}
\label{tab:wd_mass_radii}
\end{table}

\subsection{Light curve models}
\label{sec:high_speed}
\begin{figure}
  \centering
  \resizebox{1.0\columnwidth}{!}
   {
	\includegraphics{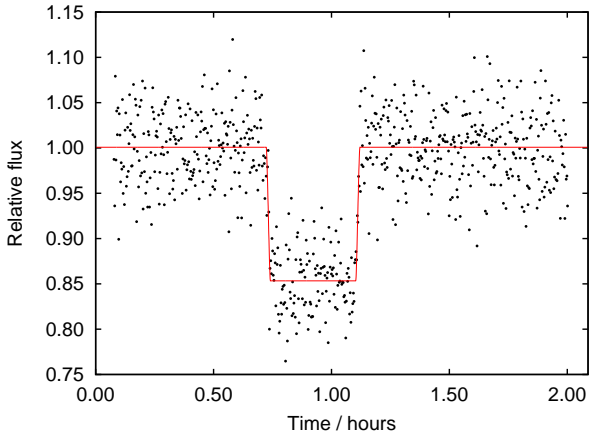}
   }
   \caption{10-second-cadence observations of PTFEB11.441 from the BOS telescope (black points). A clear filter was used to increase the signal from the target. The best-fit eclipse model is shown in red.
}

   \label{fig:high_cadence}
\end{figure}

We obtain a direct upper limit for the white dwarf radius in the PTFEB11.441 system from high-cadence eclipse data acquired with the BOS telescope (figure \ref{fig:high_cadence}). Using a clear filter to boost the signal levels, we were able to obtain 10-second-cadence data with $\approx$5\% photometric stability, sufficient time resolution to give stringent constraints on the ingress and egress timing of the system.

We fit the light curves with a simple two-spherical-body model. The flat-bottomed light curves show that the white dwarf is completely eclipsed, allowing a simple model based on the \citet{Mandel2002} transiting-planet model to be used. We modified the model to allow both stars to be moving relative to a common center of mass, to allow both bodies to be self-luminous, and to include a linear limb-darkening term for the white dwarf.

Inspection of the light curves shows that the white dwarf ingress and egress is at best marginally resolved, and so we only attempt to estimate an upper limit for the white dwarf radius. We obtain a maximal estimate of the white dwarf radius when the inclination is assumed to be $90^{\circ}$, and so we fix the inclination to that value in our modelling.  For the white dwarf, we conservatively use a linear limb-darkening coefficient of 0.5; we tested a range of coefficients and found that the limb darkening has no significant effect on the radius estimates, because the eclipse ingress and egress are barely resolved. We fix the M-dwarf radius and the stellar mass ratio $q$ during the fits, although we marginalize over the estimated 1-sigma ranges of both parameters. 

We vary the semimajor axis of the system, the white dwarf radius, and the flux from both components (and hence the contribution from the system's unresolved wide companion). When the inclination is fixed at 90 degrees, the best fit semimajor axis value is within the errorbars of our measurements based on the eclipse timing and the estimated M-dwarf radius, although those estimates suggest a somewhat lower inclination may be preferable (which would lead to a smaller white dwarf radius). We estimate the range of fitted parameters and a 95\% upper limit on the white dwarf radius using the Bootstrap algorithm.

This procedure puts the upper-limit white-dwarf radius in the PTFEB11.441 system as $<$0.025$\rm{R_{\odot}}$ at 95\% confidence. This value admits the radii estimated from cooling models and log g fits to the white-dwarf's Blamer lines. The fitted semimajor axes are around 10\% larger than those expected from the period of the system and the estimated masses of the components, suggesting we have slightly overestimated the mass of at least one component, most likely the M-dwarf in the system. 

\subsection{M-dwarf Activity}
All three systems show evidence of active M-dwarfs, including strong H$\alpha$ emission and variability on a variety of timescales. None are detected in the ROSAT All-Sky Faint Source Catalog \citep{Voges2000}, although the M-dwarfs would have to be exceptionally active to be detected at these distances (e.g. \citealt{Law2008}). 

The positioning of PTFEB11.441 near M31 allows more archival data to be found, and we identify the system with the Chandra X-ray source\footnote{from the Chandra Source Catalog \citep{Evans2010}} CXO J004545.8+415029 ($2.8\times10^{-13}\rm erg/s/cm^2$) located within 1$\sigma$ of the PTF position for the system.

\begin{figure}
  \centering
  \resizebox{1.0\columnwidth}{!}
   {
	\includegraphics{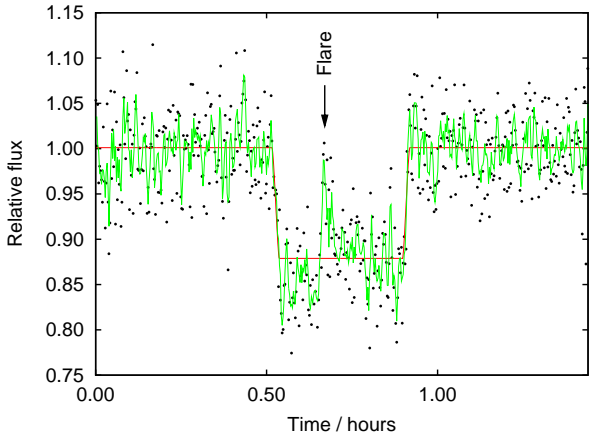}
   }
   \caption{A flare event during a high-cadence observation of PTFEB11.441. The green line shows a 30-second moving average of the 10-second-cadence photometry (black points). The red line shows the best fit eclipse light curve.}
   \label{fig:lc_flare}
\end{figure}

We also identified a flare event during one high-cadence observation of PTFEB11.441 (figure \ref{fig:lc_flare}). The observation was taken without a filter to boost our signal-to-noise, limiting the detailed interpretation of the event. The shape of the event is as expected for an M-dwarf flare, with a rapid raise followed by an approximately exponential decay.

\subsection{Out-of-eclipse variations}

The light curves show out-of-eclipse sinusoidal variability correlated
with the orbit. The overall shape is consistent with irradiation of the side of the M-dwarf facing the white dwarf, where the flux
maximum is expected half an orbital phase from the occultation. Other
mechanisms that may affect the out-of-eclipse light curves of such
compact systems are tidal ellipsoidal distortion and the beaming
effect (e.g., \citealt{Zucker2007, Shporer2010}), but simple estimates show that both effects are expected to be at the 1\% level or below for the binaries presented here and so can not be identified in the available data.

PTFEB11.441 shows much lower levels of out-of-eclipse variations than the other two systems. This is most likely due to light from its possible companion diluting the variations (and eclipse depths) in unresolved photometry of the system.

\section{Discussion and conclusions}
\label{sec:discussion}

\begin{figure}
  \centering
  \resizebox{1.0\columnwidth}{!}
   {
	\includegraphics{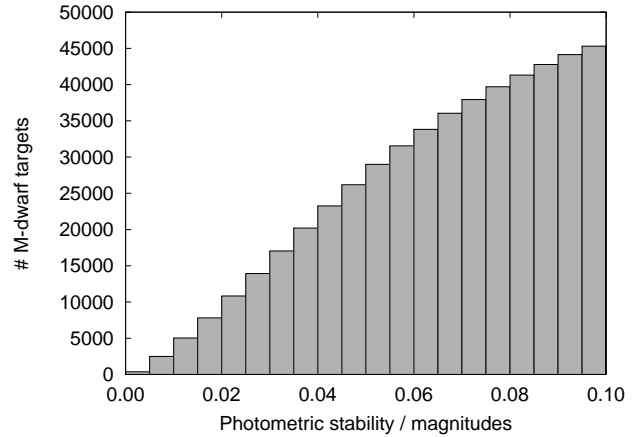}
   }
   \caption{The cumulative number of M-dwarf targets searched by the PTF/M-dwarfs survey in the first year of operations, as a function of achieved photometric stability. M-dwarfs in the PTF/M-dwarf fields are selected and confirmed on the basis of 2MASS, USNO-B1 and (where available) SDSS colors, along with proper motions determined from 2MASS and USNO-B1 positions. We require a high-confidence photometric color fit, along with a 2$\sigma$ proper motion detection to rule out giant interlopers.}
   \label{fig:n_targets}
\end{figure}

\begin{figure}
  \centering
  \resizebox{1.0\columnwidth}{!}
   {
	\includegraphics{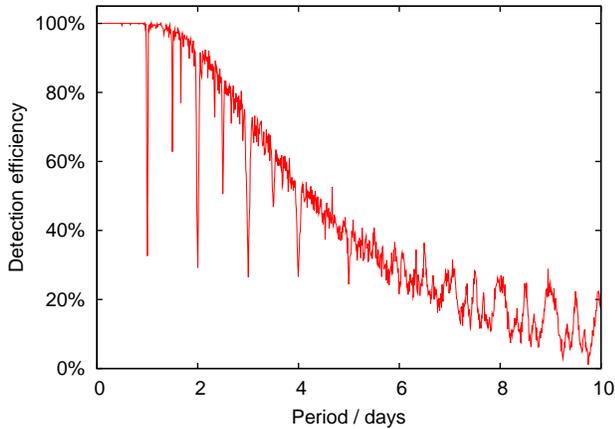}
   }
   \caption{The eclipse detection efficiency for a typical PTF/M-dwarfs observing scenario, based on simulations including the actual observing cadences of the survey. The detection efficiency is the probability that at least two eclipses will be observed from an eclipsing target with a specified period, after observing window and weather effects have been taken into account. The low efficiencies around integer numbers of days at short periods are due to the likelihood of an eclipse always occurring in daytime; the same effect allows some long period  binaries to be picked up if their eclipses always occur at night.}
   \label{fig:detec_effic}
\end{figure}

The discovery of only three white-dwarf / M-dwarf binaries in the PTF/M-dwarfs survey suggests a low incidence for such systems. This phase of the survey has covered $\approx$45,000 M-dwarfs with sufficient precision to detect short-period systems with similar eclipse depths to those shown here (figure \ref{fig:n_targets}). The survey's detection efficiency for systems with fractional-day periods is near 100\% (figure \ref{fig:detec_effic}). The geometric probability of eclipse is 5-15\% for an early-M-dwarf primary in a roughly equal-mass half-day-period binary with a white dwarf. Taken together, the detection of three systems implies that $\rm 0.08\%^{+0.10\%}_{-0.05\%}$ (90\% confidence) of M-dwarfs are in short-period post-common-envelope white-dwarf / M-dwarf binaries where the optical light is dominated by the M-dwarf.

\begin{figure}
  \centering
  \resizebox{1.0\columnwidth}{!}
   {
	\includegraphics{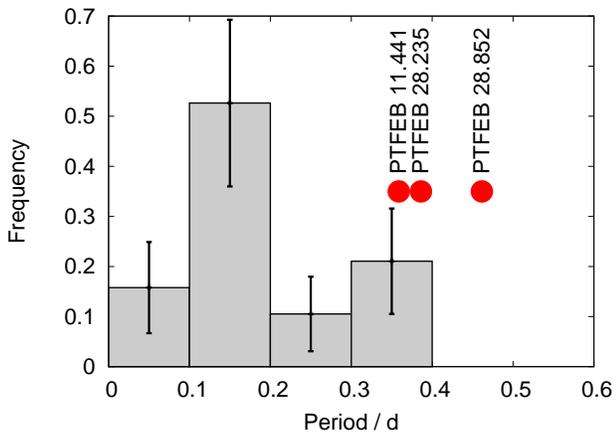}
   }
   \caption{The periods of our three detected systems compared to the distribution of periods of the white-dwarf dominated eclipsing systems found by \citet{Drake2010}, which were all found in a single survey and so offer a useful comparison sample.}
   \label{fig:period_dist}
\end{figure}

The periods of the detected systems are all longer than most known eclipsing M-dwarf / white-dwarf systems (figure \ref{fig:period_dist}). Our survey is designed to detect longer-period transiting exoplanets, so this is not in itself surprising. However, our lack of detections at shorter periods, despite our near-100\% detection efficiency for such systems, suggests that binaries including these relatively low-temperature white dwarfs are preferentially found at relatively large orbital radii.

These eclipsing binaries appear as M-dwarfs when selected by optical and infrared colors, and they display transit-like light curves which can have arbitrarily small depths. They appear to be at least three times more common than transiting giant planets in the PTF/M-dwarfs survey. These properties make these systems an important false-positive for current and future transiting planet searches around M-dwarfs. Fortunately, they can be distinguished from true transiting planets with small amounts of extra data, using the methods detailed below. The methods are ordered by difficulty, starting with those which require only the discovery light curves.

\begin{enumerate}
\item{Eclipse duration. The increased mass of the system compared to a M-dwarf/planet system leads to a shorter eclipse; high-SNR and high-cadence light curves are however required to distinguish this case from a lower-inclination system.}
\item{Eclipse shape. The eclipse of the white dwarf by the M-dwarf provides an exactly-flat-bottomed eclipse outside the short ingress and egress periods. With high-precision photometry this can be distinguished from a true transiting planet which will show the effects of limb darkening as it passes in front of the M-dwarf.}
\item{Out-of-eclipse variability. All our systems show out-of-eclipse variability at levels which are orders of magnitude greater than that expected for planetary systems (see \citet{Drake2003} for details of a similar selection method for planetary transits around solar-type stars).}
\item{UV emission. Two of our new systems showed UV emission and the third has possible emission. Where data is available, the presence of UV (or even u-band) emission suggests the existence of something other than an M-dwarf in the system.}
\item{Low-resolution spectroscopy. Like the three targets presented in this paper, a low-resolution spectrum could be decomposed into white dwarf and M-dwarf components. Faint, low-temperature white dwarfs may be, however, hard to detect with this technique.}
\item{Multi-color photometry. These systems show a strong variation in eclipse depth with wavelength (depending on the relative temperatures of the white dwarf and M-dwarf and their ratio of radii).}
\item{Radial velocities. These systems have radial velocity amplitudes at least 500$\times$ larger than expected for a planet with the same period. Just two RV few-km/sec-precision datapoints are sufficient to discriminate these systems from transiting planet systems.} 
\end{enumerate}

Of these methods, mulitcolor photometry through eclipse is probably the most time-efficient method of detecting systems like this. As large scale transiting planet surveys of M-dwarfs start up and continue, many more systems in this interesting parameter space are likely to be found.

Follow-up precision photometry and radial velocities will allow direct measurements of the masses and radii of all components of these systems. As the white dwarf transits across the disk of the M-dwarf the transit depth is expected to be around 1 millimag, although lensing by the white dwarf (e.g., \citealt{Marsh2001}) will make the transit shallower than expected from only geometric considerations (e.g., \citealt{Steinfadt2010}). High-cadence and high-precision photometry of the white dwarf eclipse (occultation) ingress and egress may be the best approach to directly measure the white dwarf radii. Furthermore, small asymmetries in the ingress and egress light curve due to the photometric Rossiter-McLaughlin effect \citep{Shporer2012, Groot2011} can allow a measurement of the white dwarf spin-orbit alignment and rotation velocity. With these methods, this new group of systems will fill a poorly-covered range of the white-dwarf and M-dwarf mass/radius relations.

\acknowledgments 
\section*{Acknowledgements}
N.M.L. is supported by a Dunlap Fellowship at University of Toronto. We thank Michael Kandrashoff, Jieun Choi and Peter Blanchard for observations at Lick Observatory.  Observations were obtained with the Samuel Oschin Telescope and the 60-inch Telescope at the Palomar Observatory as part of the Palomar Transient Factory project, a scientific collaboration between the California Institute of Technology, Columbia University, Las Cumbres Observatory, the Lawrence Berkeley National Laboratory, the National Energy Research Scientific Computing Center, the University of Oxford, and the Weizmann Institute of Science. The Byrne Observatory at Sedgwick (BOS) is operated by the Las Cumbres Observatory Global Telescope Network and is located at the Sedgwick Reserve, a part of the University of California Natural Reserve System. This paper uses observations obtained with the FTN observatory of the Las Cumbres Observatory Global Telescope. ALK was supported by NASA through Hubble Fellowship grant 51257.01 awarded by STScI, which is operated by AURA, Inc., for ANSA, under contract NAS 5-26555. The Robo-AO system is supported by collaborating partner institutions, the California Institute of Technology and the Inter-University Centre for Astronomy and Astrophysics, and by the National Science Foundation under Grant Nos. AST-0906060 and AST-0960343. AVF and his group at UC Berkeley acknowledge generous financial assistance from Gary \& Cynthia Bengier, the Richard \& Rhoda Goldman Fund, the TABASGO Foundation, and NSF grant AST-0908886. This research has also made use of the SIMBAD database, operated at CDS, Strasbourg, France. We recognize and acknowledge the very significant cultural role and reverence that the summit of Mauna Kea has always had within the indigenous Hawaiian community. We are most fortunate to have the opportunity to conduct observations from this mountain.

{\it Facilities:} \facility{Oschin, Keck, LCOGT}

\bibliographystyle{apj}
\bibliography{refs}

\label{lastpage}

\end{document}